\documentclass[conference]{IEEEtran}
\usepackage{graphics}
\usepackage{graphicx}
\usepackage{epsfig,amssymb}
\usepackage{epstopdf}
\usepackage{amsmath}
\usepackage{breqn}
\usepackage{times}
\usepackage{mathrsfs}
\usepackage{array}
\usepackage{amsmath, latexsym, amsfonts, amssymb}
\usepackage[mathscr]{eucal}
\usepackage{array, tabularx}
\usepackage[table]{xcolor}
\usepackage{multirow}
\usepackage{epsfig}

\usepackage{cite}
\usepackage[numbers,sort & compress]{natbib}

\usepackage{tikz}
\usepackage{url}
\usepackage{mathtools}
\usepackage{psfragx}
\usepackage{xcolor}
\usepackage{bbm}

\newcommand{\paren}[1]{\left(#1\right)}
\newcommand{\sqparen}[1]{\left[#1\right]}
\newcommand{\brparen}[1]{\left\{#1\right\}}

\newcommand{\defeq}{\ensuremath{\triangleq}} 
\renewcommand{\vec}[1]{\ensuremath{\boldsymbol{#1}}} 
\newcommand{\ie}{\ensuremath{{\text{\em i.e.}}}}

\newtheorem{theorem}{Theorem}
\newtheorem{lemma}{Lemma}
\newtheorem{definition}{Definition}

\IEEEoverridecommandlockouts

\begin{document}
\title{Wireless Energy Beamforming  using Received Signal Strength Indicator Feedback 
	}

\author{ Samith Abeywickrama,  Tharaka Samarasinghe, \IEEEmembership{Member, IEEE,} and Chin Keong Ho, \\ \IEEEmembership{Member, IEEE,} Chau Yuen, \IEEEmembership{Senior Member, IEEE}

\thanks{
	{The paper is accepted to be published in IEEE Transactions on Signal Processing. This is a preprint
		version. Copyright (c) 2017 IEEE. Personal use of this material is permitted. However, permission to use this material for any other purposes must be obtained from the IEEE by sending a request to pubs-permissions@ieee.org.
}
	
	{This work was supported in part by the Agency for Science, Technology and Research (A*STAR) under SERC
		Project 1420200043, and in part by the National Science Foundation of China
		(61750110529). This paper was presented in part at the IEEE Global Communications
		Conference, Washington, DC, USA, December 2016. ( \cite{wpt_gc} - Corresponding
		author: Samith Abeywickrama.)}
	
	S. Abeywickrama and C. Yuen are with the Singapore University of Technology
	and Design, Singapore (e-mail:
	\{abeywickrama\_samith,yuenchau\}@sutd.edu.sg).
	
	T. Samarasinghe is with the Department of Electronic and Telecommunication Engineering, University of Moratuwa,	Sri Lanka (e-mail: tharaka@ent.mrt.ac.lk).
	
	C. K. Ho is with the Institute for Infocomm Research, A*STAR, 
	Singapore (e-mail: hock@i2r.a-star.edu.sg).
	

	}
}

\date{}
\bibliographystyle{ieeetr}
\maketitle
\vspace{-0.5cm}
\begin{abstract}
Multiple antenna techniques that allow energy beamforming have been looked upon as a possible candidate for increasing the transfer efficiency between the energy transmitter (ET) and the energy receiver (ER) in wireless power transfer. This paper introduces a novel scheme that facilitates energy beamforming by utilizing Received Signal Strength Indicator (RSSI) values to estimate the channel. Firstly, in the training stage, the ET will transmit using each beamforming vector in a codebook, which is pre-defined using a Cramer-Rao lower bound analysis. RSSI value corresponding to each beamforming vector is fed back to the ET, and these values are used to estimate the channel through a maximum likelihood analysis. The results that are obtained are remarkably simple, requires minimal processing, and can be easily implemented. The paper also validates the analytical results numerically, as well as experimentally, and it is shown that the proposed method achieves impressive results. 
\end{abstract}

\begin{IEEEkeywords}
Wireless energy transfer, energy beamforming, received signal strength indicator (RSSI), Cramer-Rao lower bound, channel learning.
\end{IEEEkeywords}

\section{Introduction}

Wireless energy transfer (WET) focuses on delivering energy to electronic devices over the air interface.  
Electromagnetic radiation in the radio frequency (RF) bands allows us to charge freely located devices simultaneously \cite{mimo_ck}. 
When it comes to RF signal enabled WET, increasing the efficiency of the energy transfer between the energy transmitter (ET) and the energy receiver (ER) is of paramount importance. Multiple antenna techniques that also enhance the range between the ET and the ERs have been looked upon as a possible candidate to satisfy this requirement \cite{mimo_csi,ltr1,tsp11,recip1,recip2,recip3,conven,one_bit}. This paper proposes a novel approach that increases the efficiency of a WET system that uses multiple antennas to facilitate the energy transfer.

To this end, multiple antennas at the ET enable focusing the transmitted energy to the ERs via beamforming. However, the coherent addition of the signals transmitted from the ET at the ER depends on the availability of channel state information (CSI), which necessitates channel estimation. The estimation process involves analog to digital conversion and baseline processing, which require significant energy \cite{channel_esti1,channel_esti2}. Under tight energy constraints and hardware limitations, such an estimation process may become infeasible at the ER. In this paper, we propose a  method which consumes less energy, but still allows {\em almost} coherent addition of the signals transmitted from the ET at the ER. Moreover, this is a channel learning method that only requires feeding back Received Signal Strength Indicator (RSSI) values from the ER to the ET. In most receivers, the RSSI values are in fact already available, and no significant signal processing is needed to obtain them. It should be noted that the coherent addition of the signals transmitted from the ET at the ER depends directly on the phase of the channels, and it is interesting that the proposed method focuses on estimating the required phase information by only using magnitude information about the channel.

Channel estimation in WET systems normally consists of two stages. The training stage, where feedback is obtained to estimate the channel, and the wireless power beamforming (WPB) stage, where the actual WET happens. It is well known that the ET should have some knowledge (perfect or partial) about the channel to make the beamforming process productive. To this end, several methodologies for channel estimation that can be utilized for WPB have been proposed in the literature \cite{recip1,recip2,recip3,conven,one_bit,rssi_work}. In \cite{recip1,recip2,recip3}, by exploiting the channel reciprocity, the ET determines the forward link CSI by estimating the reverse link channel based on the signals transmitted by the ER. Being different to our work, these methods are mainly applicable for time division duplex (TDD) systems that use the same frequency for the uplink and the downlink. Also, using channel reciprocity for channel estimation leads to many practical difficulties, due to the non-symmetric characteristics of the RF front-end circuitry at the receiver and the transmitter \cite{reciprocity_practical}.

In \cite{conven}, the ER estimates the MIMO channel between the ET and the ER, and sends the estimated channel back to the ET. This method adopts the conventional channel estimation approach used in transmit beamforming, and it is not feasible for an ER having tight energy constraints and hardware limitations. The authors of \cite{one_bit} sought to estimate the channel using a one-bit feedback algorithm. In the training stage, the ER broadcasts a single bit to the ET indicating whether the current received energy level is higher or lower than the previous, and the ET makes phase perturbations based on the feedback of the ER to obtain a satisfactory beamforming vector for the WPB stage. This means that by utilizing the feedback bits, the ET fine tunes its transmit beamforming vector, and obtains a more refined estimate of the channel. \cite{rssi_work} could be considered to be the most related work to our work and it proposes the following methodology. In the training stage, firstly, each antenna is individually activated, and then, antennas are pairwise activated. The respective RSSI value for each activation is fed back by the ER to the ET. Next,  they utilize the gathered RSSI values to estimate the channel. 


Our proposed scheme is significantly different to \cite{recip1,recip2,recip3,conven,one_bit,rssi_work}, and our contributions and the paper organization can be summarized as follows. We focus on a system consisting of $K$ antennas at the ET, and a single antenna at the ER. We start the analysis by assuming $K=2$. Under this assumption, the proposed training stage consists of $N$ time slots. In each time slot, the ET will transmit using a beamforming vector from a pre-defined codebook of size $N$. The ER feeds back the analog RSSI value corresponding to each beamforming vector, {\em i.e.,} the ET will receive $N$ RSSI feedback values at the end of the training stage. These $N$ feedback values are utilized to set the beamforming vector for the WPB stage. More precisely, the feedback values are utilized to estimate the phase difference of the two channels between the ET and ER, and this estimate is utilized in the WPB stage. To this end, the ET equally splits the power
among the transmit antennas, and pre-compensates channel
phase shifts such that the signals are coherently added up at
the ER regardless of the channel magnitudes. These ideas are introduced in Section \ref{Section:System model}. 

In Section \ref{section:crlb}, we focus on defining the aforementioned pre-defined codebook. To this end, we employ a Cramer-Rao lower bound (CRLB) analysis, and define the codebook such that the estimator of the phase difference between the two channels of interest achieves the CRLB, which is the best performance that an unbiased estimator can achieve. On top of providing a solid theoretical basis for the selection of the beamforming vectors for the training stage, this approach also allows us to simplify derived  results significantly, and most importantly, it leads to achieving  impressive results in the WET. The defined codebook gives the ET sufficient  information to obtain the $N$ RSSI feedback values. In Section \ref{Section:Channel Estimation}, we discuss how the feedback values can be utilized to set the beamforming vector for the WPB stage, through a maximum likelihood analysis. Our analysis takes the effect of noise on the measurements into account unlike \cite{rssi_work}. The results that we obtain are remarkably simple, requires minimal processing, and can be easily implemented at the ET. Also, the results are general such that they will hold for all well known fading models. However, it should be noted that the estimate, which is a phase value, has an ambiguity due to the use of $ \tan^{-1} $, and hence can take two values. In \cite{rssi_work}, a similar phase ambiguity is resolved by ascertaining further RSSI feedback (four values) for the candidate phase values from the ER, and picking the candidate that provides the best energy transfer. In this paper, we propose a  method, that allows us to resolve the ambiguity without ascertaining any further RSSI feedback from the receiver.


In section \ref{section: N large}, we show how our results can be extended for a single-user WET system consisting of $ K > 2 $ antennas at the ET. Then, we focus on selecting $N$. Although larger $ N $ yields a higher channel estimation precision, for a given time period $ T $, a larger $ N $ will consume a larger portion of $ T $, which will reduce the time for WPB. Therefore, larger $N$ may lead to a reduction in the total transferred energy. In Section \ref{opt_N}, we present  bounds for the optimal value of $N$ that maximizes the system performance in terms of the energy transfer during the WPB stage. In Section \ref{section:Numerical Evaluations}, we validate our analytical results numerically, while providing useful insights into the system performance. Furthermore, Section \ref{Experimental Validation} shows that the proposed methodology can be in fact implemented on hardware, and the experimental setup is used to further validate our results. Experimental validation is not common in the related works, and can be highlighted as another major contribution of this paper. Both Section \ref{section:Numerical Evaluations} and Section \ref{Experimental Validation} show that our proposed method will achieve impressive results, and will provide performance improvements compared to directly related works in the literature. It should be also noted that the proposed methodology can be used for any application of beamforming in which processing capabilities of the receiver are limited. Section \ref{Section:Conclusions} concludes the paper.

\begin{figure}[t]
	\centering {\includegraphics[scale=1]{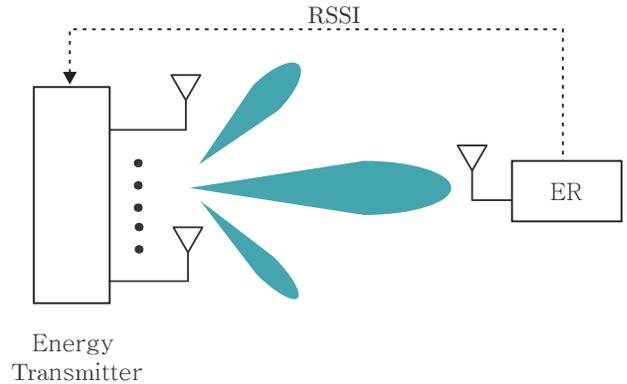}} 
	\caption{System model.}	 
	\label{sm}
\end{figure}

\section{System Model and Problem Setup}\label{Section:System model}

We consider a MISO channel for WET. An ET consisting of $K \geq 2$ antennas delivers energy to an ER consisting of a single antenna over a wireless medium, see Fig. \ref{sm}.
The transmit signal at the ET is given by $ \vec{x} = \vec{w} s $, where $\vec w \in \mathbb{C}^{K \times 1}$ denotes the complex $K$-by-$1$ beamforming vector and $ s $ denotes the transmit symbol, which is independent of $\vec{w}$, and has zero-mean and unit variance (\textit{i.e.,} $ \mathbb{E}( |s|^{2}) =1 $). We have dropped the time index for notational simplicity. The transmit covariance  matrix is given by $  \mathrm {\bf C}_{\vec {xx}} = \mathbb E (\vec{x}\vec{x}^{\dag}) = \mathbb E  (\vec{w}\vec{w}^{\dag})$, where $ \dag $ denotes the conjugate transpose. $ \mathrm {\bf C}_{\vec {xx}} $ is positive semi-definite, thus the number of energy beams $ d $ can be obtained from the 
rank of $ \mathrm {\bf C}_{\vec {xx}} $ \cite{power1}, \textit{i.e.,} $ d= \mathrm{rank}(\mathrm {\bf C}_{\vec {xx}}) $ . It is assumed that the maximum transmit sum-power constraint at the ET is $ P > 0 $.  Therefore, we have $ \mathbb{E} (\|\vec{x}\|^{2}) = \mathrm{tr}(\mathrm {\bf C}_{\vec {xx}}) \leq P$, where $ \mathrm{tr}(\cdot) $ denotes the trace of a square matrix, and $ \|\cdot \| $ denotes the Euclidean norm. 

Let $ \vec{h} = \sqparen {|h_{1}|e^{j\delta_{1}}, \dots , |h_{K}|e^{j\delta_{K}}}^\top  $ represent the complex MISO channel vector between the ET and the ER. Further, we consider a quasi-static block-fading channel model and a block-based energy transmission, where it is assumed that the wireless channel remains constant over each transmission block. The transmission block has a length $ T > 0 $ (in practice, $T$ is upper bounded by the channel coherence time).  The received energy (or RSSI) at the ER can be written as
\begin{eqnarray}
\mathrm R = \xi (\vec h^{\dag} \mathrm {\bf C}_{\vec {xx}}\vec h), \label{power}
\end{eqnarray}
where $\xi$ denotes the conversion efficiency of the energy harvester \cite{power1}. 

Our main focus is to design a \textit{single energy beam} to maximize the received energy at the ER, so that the harvested energy is maximized at the ER. To this end, we focus on the following optimization problem:
\begin{equation}
\begin{aligned}
& \underset{\footnotesize \mathrm {\bf C}_{\vec {xx}} \succeq 0}{\text{maximize}}
& &\xi (\vec h^{\dag} \mathrm {\bf C}_{\vec {xx}}\vec h) \nonumber \\
& \text{subject to}
& & \mathrm{tr}(\mathrm {\bf C}_{\vec {xx}}) \leq P , \enspace \mathrm{rank}(\mathrm {\bf C}_{\vec {xx}}) =1.
\end{aligned}
\end{equation}
Since $ \mathrm{rank}(\mathrm {\bf C}_{\vec {xx}}) = 1 $, $ d=1 $. The solution for this optimization  problem  is  $ \mathrm {\bf C}_{\vec x \vec x}^\star = P\vec{v}\vec{v}^{\dag} $, where $ \vec v $ denotes the dominant eigenvector of the normalized MISO channel covariance matrix $ \mathrm {\bf H} $ \cite{power1}, \textit{i.e.}, $ \mathrm {\bf H} =  \frac{\vec{h}\vec{h}^{ \dag}}{\|\vec{h}\vec{h}^{ \dag}\|_{\mathrm F}}     $, and $  \| \cdot \|_{\mathrm F} $ denotes the Frobenius norm. 

We employ equal gain transmit (EGT) beamforming for the WET. Thus, the optimal transmit signal can be written as $ \vec{x}^\star = \sqrt{P} \vec{v} s$, which implies an optimal beamforming vector $ \vec{w}^\star = \sqrt{P} \vec{v}  $. To this end, 
\begin{dmath}
	\displaystyle	\vec{v} = \frac{1}{\sqrt{K}} \sqparen {1, e^{-j\phi_{2}} , \dots ,e^{-j\phi_{K}} }^\top,
	\label{vec}
\end{dmath}
where $ \phi_{k} = \delta_{k} - \delta_{1} $, $ k \in \{2,\dots,K\} $. In practice, each transmit antenna has its own power amplifier, which operates properly only when the transmit power is below a pre-designed threshold. Therefore, there are practical difficulties in implementing maximum ratio transmit (MRT) beamforming, where the transmit power in some antennas may theoretically exceed these threshold values. Because of this reason, although MRT is superior, still, EGT beamforming, where the ET equally splits the power among all transmit antennas, is a preferred method in practice \cite{why_mgt}. In this paper, we assume that the pre-designed transmit power threshold is equal among antennas, and we transmit at that power. It should be noted that our results can be easily extended to a case where these threshold values are not equal among antennas as well. More specifically, the results can be extended to general sum-power or per-antenna power constraints, but the power allocation among antennas will be static, and not dynamic as in a case where the ET employs MRT beamforming.

From \eqref{vec}, we can see that the optimality of the wireless energy transfer depends only on $\brparen{\phi_{k}}_{k=2}^{K}$, and these values can be set without any loss of optimality if full channel state information (CSI) is available at the ET. In practice, full CSI at the ET can be achieved by estimating the channel at the ER, and feeding back the channel information to the ET. 
However, we are particularly focusing on applications with tight energy constraints at the ER. Thus, such an estimation process may become infeasible as channel estimation involves analog to digital conversion and baseline processing, which require significant energy. Therefore, we focus on introducing a more energy friendly  method of selecting the beamforming vector, by only considering RSSI values that are fed back from the ER to the ET. It should be noted that the feedback takes the form of real values, and RSSI values are readily available in most receiver circuits. We will first present the proposed scheme for the special case of $ K=2 $ to draw useful insights, and then, in Section \ref{section: N large}, we will extend the proposed scheme to the general case of $ K> 2 $. 

\begin{figure}[t]
	\centering {\includegraphics[scale=0.75]{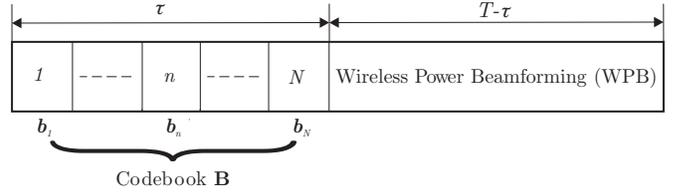}} 
	\caption{The two-phase transmission protocol when $ K=2 $.}	 
	\label{2pha_1}
\end{figure}

Under the assumption of $K=2$, the proposed scheme is as follows. The scheme consists of a training stage and a wireless power beamforming (WPB) stage. As we have depicted in Fig. \ref{2pha_1}, the training stage is further divided into $ N$ mini slots. We define a codebook $ \mathrm {\bf{B}} = \sqparen{\vec{b}_{1} \ 	\ldots \ \vec{b}_{N}} $ that includes $N$ beamforming vectors to be used in each mini slot in the training stage. Let $ \mathcal{N} = \{1,\dots,N\}$. In each mini slot $n \in \mathcal{N}$, the ET simultaneously activates both its antennas and transmits using beamforming vector $ \vec{b}_{n} \in \mathrm {\bf{B}}$. This means, the $ n $-th element of $ \mathrm {\bf{B}} $ is used in $ n $-th mini slot. Let $ \mathrm R_{n} $ denote the RSSI value at the ER during mini slot $n \in \mathcal{N}$. The ER will feedback $\mathcal{R}=\brparen{\mathrm R_{n}}_{n=1}^N$ to the ET, which means, at the end of the training period $\tau$, the ET will have $N$ RSSI feedback values corresponding to each element in $\mathrm {\bf{B}}$.
  
Moreover, consider the $n$th element of $\mathrm {\bf{B}}$ to take the form of $\vec{b}_{n} = \sqrt{\frac{P}{2}} \sqparen {1 \enspace e^{j\theta_{n}} }^\top $, where $\theta_{n}$ is the $n$-th element of $\Theta$. $\Theta$ is a set that includes phase values between $ 0 $ and $ 2\pi $. For implementation convenience, $\Theta$ is predetermined and does not depend on the feedback values. Further,  we shall employ estimation theory and the concept of the CRLB in order to define $\Theta$. At the end of the training stage, the ET will determine the beamforming vector $\vec{w}_{\text{WPB}} $ to be used for the WPB stage. The ER does not feed back in the WPB stage, and typically, this stage is longer than the training stage to reduce the overhead incurred in the WPB stage. From \eqref{vec}, it is not hard to see that the optimal beamforming vector should take the form of $\vec{w}_{\text{WPB}} = \sqrt {\frac{P}{2}} \sqparen {1 \enspace e^{-j\phi_{2}} }^\top$.  Our challenge is to estimate ${\phi_2}$ by only utilizing $\mathcal{R}$.  
  
We denote the RSSI value at the ER during mini slot $n \in \mathcal{N}$ by $R_n$, and it is written as
\begin{equation}
\mathrm R_{n}= \xi (\vec h^{\dag} \mathrm {\bf C}_{\vec {xx}}\vec h) +  z_{n}. 
\label{eq:ch rssi}
\end{equation} 
Note that due to noise, the RSSI value will change from one mini slot to the other. We use random variable $ z_{n}$ to capture the effect of noise on $  \mathrm R_{n} $.  More specifically,  $ z_{n} $ captures the effect of  all noise related to the measurement process such as noise in the channel, circuit, antenna matching network and rectifier. We assume that the channel is slowly varying so that during the training stage and the subsequent beamforming, $ \vec{h} $ can be considered to be unknown, but non varying (fixed). Therefore, the randomness in \eqref{eq:ch rssi} is caused only by  $z_{n}$. For tractability, and without loss of generality, we assume $\vec{z}= \sqparen{z_{1},\ldots, z_{N}}^\top $ to be an i.i.d. Gaussian random vector with zero mean and variance $\sigma^2$. 

Under the above assumptions, for $K=2$, $\ie$,  a channel vector $ \vec{h} = \sqparen {|h_{1}|e^{j\delta_{1}} \ \ |h_{2}|e^{j\delta_{2}}}^{\top}$. we have 
\begin{equation*}
\mathrm {\bf C}_{\vec x \vec x}
= \frac{1}{2}
\begin{bmatrix}
1 &  e^{j\theta_n}\\
e^{-j\theta_n} &  1\\
\end{bmatrix}.
\end{equation*}
Thus, \eqref{eq:ch rssi} can be simplified as
\begin{alignat} {2}
\mathrm R_{n} 
&= \frac{\xi P}{4} \Big(|h_{1}|^{2} + |h_{2}|^{2} + 2|h_{1}||h_{2}|\cos\paren{\theta_{n}+\delta_{2}-\delta_{1}}\Big) + z_{n} \nonumber\\
&=\alpha + \beta \cos\paren{\theta_{n}+\phi_{2}} + z_{n},
\label{eq:ch rssi_2}
\end{alignat} 
where $\alpha = \frac{\xi P}{4}(|h_{1}|^{2} + |h_{2}|^{2})$, $\beta = \frac{\xi P}{2}|h_{1}||h_{2}|$, and $\phi_{2}= \delta_{2}-\delta_{1}$. Our goal is to estimate $\phi_2  $. It can be seen from \eqref{eq:ch rssi_2} that $  \mathrm R_{n} $ depends on three unknown parameters $ \alpha$, $\beta$, and $ \phi_2 $. Hence, the parameter vector can be written as $\vec \varphi = [\alpha \enspace \beta \enspace \phi_2]^{\top} $. 

To implement the proposed method in this paper, we should first define $\Theta$. In the next section, we define $\Theta$ by performing a CRLB analysis on the parameter vector. Then, $\Theta$ will be used to define the codebook $\mathrm {\bf{B}}$, and in Section \ref{Section:Channel Estimation}, we discuss how the RSSI feedback values associated to the beamforming vectors in $\mathrm {\bf{B}}$ can be used to estimate $\phi_2$ through a maximum likelihood analysis.

\section{Cramer-Rao lower bound analysis } \label{section:crlb}

The CRLB is directly related to the accuracy of an estimation process. More precisely, the CRLB gives a lower bound on the variance of an unbiased estimator. To this end, suppose we wish to estimate the parameter vector  
$ \vec {\varphi} = [\alpha \enspace \beta \enspace \phi_2]^{\top} $. The unbiased estimator of $ \vec {\varphi} $ is denoted by $ \hat {\vec {\varphi}} = [\hat\alpha \enspace \hat\beta \enspace \hat\phi_2]^{\top} $, where $ \mathbb E \{\hat {\vec {\varphi}} \} = \vec {\varphi}$. The variance of the unbiased estimator $ \mathrm{var}( {\hat{\vec \varphi}}) $ is lower-bounded by the CRLB of $ \vec {\varphi} $, which is denoted by $ \mathrm{CRLB_{\vec {\varphi}}} $, \textit{i.e.},
$	\mathrm{var}(\hat{\vec \varphi}) \geqslant \mathrm{CRLB_{\vec {\varphi}}} $.
Moreover, $\mathrm{CRLB_{\vec {\varphi}}} $ can be obtained by the inverse of $ \mathrm{FIM}_{\vec \varphi} $, which is the Fisher information matrix (FIM) of $\vec \varphi$.
Since no other unbiased estimator of $ \vec {\varphi} $ can achieve a variance smaller than the CRLB, the CRLB is the best performance that an unbiased estimator can achieve. Hence, our motivation is to select $ \Theta $ in a manner that the estimator achieves the CRLB, and its variance is minimized. Also note that as discussed in Appendix \ref{App:crlb}, the Gaussian distribution leads to the worst-case CRLB performance for our estimation problem. Therefore, due to the Gaussian assumption made on the random variable in \eqref{eq:ch rssi}, we are minimizing the largest or the worst case CRLB. 

Using \eqref{eq:ch rssi}, the $N$-by-$1$ vector representing $N$ RSSI observations can be written as
\begin{alignat}{2} 
\mathrm{\bf{R}}=  \vec x_{\varphi} +  \vec z,
\label{final_model}
\end{alignat}
where $\vec x_{\varphi} $ is a $N$-by-$1$ vector of which the $n$th element takes the form of $ \alpha + \beta \cos(\theta_{n} + \phi_2)$. Since $\vec x_{\varphi}$ is independent of $\vec z$, $\mathrm{\bf{R}}$ in \eqref{final_model} is distributed according to a multivariate Gaussian distribution, {\em i.e.}, 
$ \mathrm{\bf{R}} \sim \mathcal{N} (\vec x_{\varphi}, \mathrm {\bf{C} }_{zz}) $,
where $ \mathrm {\bf{C} }_{zz}= \sigma^2 \mathrm {\bf{I}_N} $, and $\mathrm {\bf{I}_N}$ is the $N$-by-$N$ identity matrix. We will specifically focus on $ \phi_2 $, which is the main parameter of interest, and derive the CRLB of its estimator. Then, we will focus on finding the set of values $\brparen{\theta_{n}}_{n=1}^N$ that will minimize the derived CRLB. We will start by deriving the FIM of $ \vec {\varphi} $, which is formally presented in the following Lemma.  
\begin{lemma} \label{Lemma: FIM}
	The FIM of $ \vec {\varphi} $ is given by
	\begin{alignat*}{2}
	\mathrm{FIM}_{\varphi}(\mathrm{\bf{R}})
	= \frac{1}{\sigma^{2}}
	\begin{bmatrix}
	  N   &   \sum_{n=1}^{N} A_{n}   &   \sum_{n=1}^{N} D_{n}   \\\\ 
	\sum_{n=1}^{N} A_{n}   &   \sum_{n=1}^{N}  A_{n}^2  &   \sum_{n=1}^{N} A_{n}D_{n}  \\\\ 
	\sum_{n=1}^{N} D_{n}   &   \sum_{n=1}^{N} A_{n}D_{n}   &   \sum_{n=1}^{N} D_{n}^2  \\
	\end{bmatrix},
	\end{alignat*}
	where $A_{n} = \cos(\theta_{n}+\phi_2)$ and $D_{n}  = -\beta \sin(\theta_{n}+\phi_2)$.
\end{lemma}
\begin{IEEEproof}
	See Appendix \ref{App:crlb}.
\end{IEEEproof}

We will first use the FIM to obtain some useful insights on the selection of $N$. These insights can be drawn from the determinant of the FIM. To this end, for $ N=1 $ and $ N=2 $, $ \det(\mathrm{FIM}_{\varphi}(\mathrm{\bf{R}})) = 0 $, which implies that $ \mathrm{FIM}_{\varphi}(\mathrm{\bf{R}}) $ is not invertible for these two cases. Since the CRLB of $ \phi_2 $ is the 3rd diagonal element of the inverse of $ \mathrm{FIM}_{\varphi}(\mathrm{\bf{R}})  $, we can conclude that the CRLB is unbounded when $ N<3 $. Therefore, the estimation variance of $ \phi_2 $ is unbounded when $ N<3 $, implying that we need at least 3 RSSI values fed back to the ET to make the proposed scheme work. On the other hand, when $N \geq 3$, we have 
\begin{alignat*}{1}
\det(\mathrm{FIM}_{\varphi}(\mathrm{\bf{R}})) = 
\displaystyle \beta^{2}  \sum_{i=1}^{N-2}  \sum_{j=i+1}^{N-1}  \sum_{k=j+1}^{N} \Delta_{i,j,k}, 
\end{alignat*}
where
\begin{alignat*} {2}
\Delta_{i,j,k}=  \Big[4\sin \Big(\frac{\theta_{i}-\theta_{j}}{2}\Big)\sin \Big(\frac{\theta_{j}-\theta_{k}}{2}\Big)\sin \Big(\frac{\theta_{k}-\theta_{i}}{2}\Big) \Big]^{2}, \end{alignat*}
which will be non zero if $\Theta$ consists of $N$ distinct phase values\footnote{Obtaining this expression analytically is straightforwardly done by computing
the determinant of a 3-by-3 matrix. However, due to being tedious, it is
omitted to avoid any deviation from the main focus of the paper.}. Thus, if $\Theta$ is selected accordingly, the CRLB will exist for $N \geq 3$. Along these ideas, we will use $ \mathrm{FIM}_{\varphi}(\mathrm{\bf{R}}) $ to  derive the CRLB of $\phi$, and it is formally presented through the following lemma. 
\begin{lemma} \label{Lemma: CRLB of phi}
For $N\geq 3  $, if $\Theta$ consists of $N$ distinct phase values, the CRLB of parameter $ \phi_2 $ exists, and it is given by
\begin{alignat*}{1} 
\mathrm{CRLB}_{\phi} = \frac{\displaystyle \sigma^{2} \sum_{i=1}^{N-1} \sum_{j=i+1}^{N} \Big[\cos(\theta_{i}+\phi_2) - \cos(\theta_{j}+\phi_2) \Big]^{2} }  {\displaystyle \beta^{2}  \sum_{i=1}^{N-2}  \sum_{j=i+1}^{N-1}  \sum_{k=j+1}^{N} \Delta_{i,j,k}   }.
\end{alignat*}
\end{lemma}
\begin{IEEEproof}
See Appendix \ref{App:crlb}.
\end{IEEEproof}

Having derived the CRLB of $\phi_2$, our goal is to find $\brparen{\theta_{n}}_{n=1}^N$ that will minimize the derived CRLB for any given $\phi_2$. However, it should be noted that the $ \mathrm{CRLB}_{\phi} $  is a function of $\phi_2$. Therefore, the CRLB minimizing $\brparen{\theta_{n}}_{n=1}^N$ will be functions of $\phi_2$ as well. This will lead to implementation difficulties as $\Theta$ is supposed to be predefined. Therefore,  we resort to averaging out the effect of $\phi_2$. To this end, we assume $\phi_2$ to be uniformly distributed in $(0, 2\pi]$, and computing the expectation over $\phi_2$ leads to the \textit{modified Cramer-Rao lower bound} (MCRLB) \cite{navigation}. The MCRLB is formally presented through the following lemma, and the proof is skipped since its trivial.
\begin{lemma}
The MCRLB of parameter $ \phi_2 $ is given by
\begin{alignat}{2} 
\mathrm{MCRLB}_{\phi} &= \mathbb E_{\phi} [\mathrm{CRLB}_{\vec \phi}]  \nonumber \\ 
&= \frac{\displaystyle \sigma^{2}\sum_{i=1}^{N-1} \displaystyle \sum_{j=i+1}^{N} \Big[1 - \cos(\theta_{i}-\theta_{j}) \Big] }  {\displaystyle \beta^{2} \displaystyle \sum_{i=1}^{N-2} \displaystyle \sum_{j=i+1}^{N-1} \displaystyle \sum_{k=j+1}^{N} \Delta_{i,j,k}   }. 
\label{crlb2}
\end{alignat}
\label{lemma2}
\end{lemma} 

After obtaining the MCRLB, our goal shifts to finding the MCRLB minimizing $\brparen{\theta_{n}}_{n=1}^N$. Determining the MCRLB minimizing $\brparen{\theta_{n}}_{n=1}^N$ analytically for a general case is not straightforward due to the complexity of \eqref{crlb2}. To develop insights, we will first focus on the $N=3$ case and derive the MCRLB minimizing $\brparen{\theta_{1},\theta_{2},\theta_{3}}$. To this end, without any loss of generality, we assume  $\theta_{1}$ to be zero and $\theta_2$ and $\theta_3$ are set relative to $ \theta_{1} $. Then, we repeat the process for $N=4$. From these two derivations, we can observe a pattern in the $ \mathrm{MCRLB}_{\phi} $ minimizing $\theta_{n}$ values, and we define $\Theta$ by making use of this pattern. In Section \ref{section:Numerical Evaluations}, through numerical evaluations, we validate the selection of $\Theta$ for arbitrary values of $N$. 

\begin{lemma}
Let $\theta_1=0$. For $N=3$, $\Theta=  \brparen{0 , 2\pi /3, 4\pi/3 }$ minimizes $ \mathrm{MCRLB}_{\phi} $, and the corresponding minimum value is $\frac{2 \sigma^{2}}{3 \beta^{2}}$. For $N=4$, $\Theta=  \brparen{0 , \pi/2,  \pi, 3\pi/2 }$ minimizes $ \mathrm{MCRLB}_{\phi} $, and the corresponding minimum value is $\frac{2 \sigma^{2}}{4 \beta^{2}}$. 
\label{lemma3}
\end{lemma}
\begin{IEEEproof}
See Appendix \ref{App:crlb}.
\end{IEEEproof}

It is interesting to note that in both cases, the phase values in $\Theta$ are equally spaced over $ [0 \enspace 2\pi) $. For an example, when $N=3$, $ |\theta_{1} - \theta_{2}| = |\theta_{2} - \theta_{3}| = |\theta_{3} - \theta_{1}| = 2\pi/3$. When $N=4$, the phase difference between adjacent elements in the set turns out to be  $2\pi/4$. Also, by observing this pattern,we can expect  the minimum $ \mathrm{MCRLB}_{\phi} $ to behave like  $\frac{2 \sigma^{2}}{N \beta^{2}}$ with $N$. To this end, we will define $\Theta$ for $N$ elements as follows.
\begin{definition} \label{Def: Def1}
$\Theta$ is a set of phase values between $0$ and $2\pi$, and it is defined to be $\Theta= \brparen{\theta_1, \ldots, \theta_N}$, where $\theta_n= \frac{2(n-1)\pi}{N}$ for $n \in \mathcal{N}$. 
\end{definition}

\begin{figure}[t]
 {\includegraphics[trim = 20mm 0mm 0mm 0mm, clip, scale=0.5]{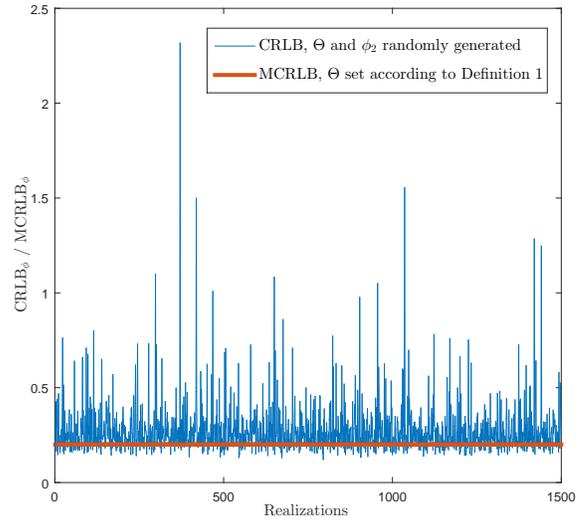}} 
	\caption{Comparison between CRLB and MCRLB.}	 
	\label{CRLBvsMCRLB}
\end{figure}

The intuition behind this definition is that getting RSSI values with the maximum spatial diversity provides us the best estimate. Using the phase values in $\Theta$, $N$ RSSI feedback values can be obtained. It should be stressed that although our initial goal was minimizing the CRLB, we ended up minimizing the MCRLB, which is obviously not the same thing. As shown in \cite{avg_crlb}, sometimes, depending on the averaging, minimizing the MCRLB might lead to inferior results. Therefore, in order to check the effectiveness of the MCRLB for our application, a simple test was carried out, and the results are illustrated in Fig. \ref{CRLBvsMCRLB}. In this test, for a predetermined $N$, we randomly generated $\brparen{\theta_n}_{n=1}^N$ and $\phi_2$ assuming they are uniformly distributed between 0 and 2$\pi$, and evaluated the CRLB in Lemma \ref{Lemma: CRLB of phi}. As shown in the figure, this was done for 1500 realizations. Then, we evaluated the MCRLB for the same $N$, and $\brparen{\theta_n}_{n=1}^N$ selected according to $\Theta$ defined in Definition \ref{Def: Def1}. The comparison is presented in Fig. \ref{CRLBvsMCRLB}, and it can be seen that the MCRLB with $\Theta$ defined according to Definition \ref{Def: Def1} is a very reasonable approximation for the lower bound of the CRLB. Therefore, although minimizing the MCRLB instead of the CRLB is suboptimal, the loss of optimality is negligibly small. 

Having obtained the feedback values at the ET, the next question is how these feedback values can be used to estimate the phase difference between the two channels. This, question is addressed in the next section.

\section{Estimation of the Channel Phase Difference $ \phi_2 $} \label{Section:Channel Estimation}

\subsection{Estimating $ \phi_2 $ in a Noiseless Environment}

We will first look at a simplified scenario similar to \cite{rssi_work} by neglecting the effect of noise. If there is no noise in the network, we have $ \mathrm{R}_{n}= \alpha + \beta \cos{(\theta_{n}+\phi_2)} $ for $n \in \mathcal{N}$. If $ N=3 $, we can simply calculate $\phi_2$ by solving three simultaneous equations, after obtaining three RSSI feedback values. 
The result is formally presented in the following theorem and this value of $\phi_2$ should intuitively give satisfactory results in low noise environments. The proof is skipped as it is trivial.  
\begin{theorem} \label{th_no_noise}
In a noiseless environment, for $ N=3 $, and $ \Theta $ defined according to Definition 1, the estimate of the phase difference between the two channels between the ET and the ER is given by
\begin{alignat}{2}
\hat\phi_2 &= \tan^{-1}\paren{\frac{  \sqrt{3} \lambda_{2,3} }          
	{  \lambda_{2,1} + \lambda_{3,1}
	}	
},
\raisetag{-.5em}
\end{alignat}
where $ \lambda_{i,j} = \mathrm {R}_{i} - \mathrm {R}_{j}$ for $i,j \in$ \{1,2,3\}.
\end{theorem}

It should be noted that $\hat \phi_2 $ has an ambiguity due to the use of $ \tan^{-1} $, and $\hat \phi_2 $ can take two values $\hat{\phi}_{2,1}$ and $\hat{\phi}_{2,2}$, such that $\hat{\phi}_{2,2}=\hat{\phi}_{2,1}-\pi$. In \cite{rssi_work}, the ambiguity is resolved by introducing an ambiguity resolution stage, right after the training stage. In the ambiguity resolution stage, the ET sequentially beamforms using each candidate value of $\hat{\phi}_2$, and obtains the respective RSSI values from the ER through feedback. Then, the ET picks the candidate that provides the best energy transfer to set the beamforming vector for the WPB stage. It should also be noted that \cite{rssi_work} requires the ET to acquire four more feedback values to resolve the ambiguity as the phase difference is given as $ \cos^{-1} $. In this paper, we propose a more energy efficient method, that allows us to resolve the ambiguity without ascertaining any further RSSI feedback values. This will be discussed later in this section.

\subsection{Estimating $ \phi_2 $ in Noisy Environments}

Now, we will focus on the more general  scenario. Firstly, we will present the following auxiliary results, that will be directly used in the proofs of the main results.
\begin{lemma} \label{aux}
Let $ \theta_{n} = \frac{2(n-1)\pi}{N}$ for $n \in \mathcal{N}$. Then, 
\begin{multline*}
\sum_{n=1}^{N} \sin(\theta_{n}+\phi_2)=\sum_{n=1}^{N} \sin{[2(\theta_{n}+\phi_2)]}= \sum_{n=1}^{N} \cos(\theta_{n}+\phi_2) \\ =\sum_{n=1}^{N} \cos{[2(\theta_{n}+\phi_2)]}=0.
\end{multline*}
\end{lemma}
\begin{IEEEproof}
See Appendix \ref{App:Channel Estimation}.
\end{IEEEproof}

Based on the assumption that the effect of noise is i.i.d. Gaussian, estimating $\phi_2$ becomes a classical parameter estimation problem. Thus, a maximum likelihood estimate of $\phi_2$ can be obtained by finding the value of $\phi_2$ that minimizes 
\begin{eqnarray}
\mathrm E \defeq \sum_{n=1}^{N} \Big[\mathrm {R}_{n} - (\alpha + \beta \cos{(\theta_{n}+\phi_2)}) \Big]^{2}. \label{ml}
\end{eqnarray}
Differentiating $\mathrm E$ with respect to  $\phi_2$, and setting it to zero gives us
\begin{dmath}
	\sum_{n=1}^{N} \mathrm R_{n}\sin{(\theta_{n}+\phi_2)} = \alpha \sum_{n=1}^{N} \sin{(\theta_{n}+\phi_2)} + \frac{\beta}{2} \sum_{n=1}^{N} \sin{[2(\theta_{n}+\phi_2)]}.
	\label{Estimation_phi}
\end{dmath}
It is not hard to see that to obtain the solution of $\phi_2$, we have to first estimate $\alpha$ and $\beta$, and these non-essential parameters are referred to as nuisance parameters \cite{nuisance}. However, thanks to the way we have defined $\Theta$, we can obtain an ML estimate of $\phi_2$ without estimating the nuisance parameters. These ideas are formally presented in the following theorem.  
\begin{theorem}\label{esti}
	For a sample of $ N $ i.i.d. RSSI observations,  $ \phi_2 $ can be estimated by 
	\begin{gather}
	\hat\phi_2 = \tan^{-1}\paren{\frac{\displaystyle  -\sum_{n=1}^{N} \mathrm R_{n} \sin\theta_{n} }           
		{\displaystyle \sum_{n=1}^{N} \mathrm R_{n} \cos\theta_{n}}	
	},
	\label{estimation_phi}
	\end{gather}
	where $ \theta_{n} = \frac{2(n-1)\pi}{N}$ for $ n \in \mathcal{N} $.
\end{theorem}
\begin{IEEEproof}
See Appendix \ref{App:Channel Estimation}.
\end{IEEEproof}	

We can observe that $\tan(\phi_2)$ is the ratio between two weighted sums of the same set of RSSI values. The $i$-th RSSI value in the denominator is weighted by the cosine of an angle, {\em i.e.}, $\cos(\theta_i)$, where as in the numerator, the same RSSI value is weighted by the cosine of the same angle, but shifted by 90 degrees, {\em i.e.}, $\cos(\frac{\pi}{2}+\theta_i)$. Setting these angles according to Definition 1 gives us the best estimate of $\tan(\phi_2)$, which leads to the best estimate of $\phi_2$. Note that the result in Theorem \ref{esti} is easy to calculate, requires minimal processing, and can be easily implemented at the ET. 
We should stress that the simplicity of the result was mainly possible due to the CRLB analysis performed in Section \ref{section:crlb} to define $\Theta$. However, it should be noted that similar to the noiseless case, $ \phi_2 $ has an ambiguity due to the use of $ \tan^{-1} $, and next, we will discuss how this can be resolved.   

\subsection{Resolving the ambiguity  of the estimate of $ \phi_2 $} \label{ambiguity}

We propose a method of selecting the correct estimate of $ \phi_2 $ and resolving the ambiguity without ascertaining any further RSSI feedback values from the ER. This idea is formally presented through the following theorem.

\begin{theorem}\label{ambi}
	Let $\hat{\phi}_{2,1}$ and $\hat{\phi}_{2,2}$ be the possible solutions for the estimate of $\phi_2 $, where $\hat{\phi}_{2,2}=\hat{\phi}_{2,1}-\pi$. Then, the RSSI maximizing solution $\phi_2^\star$ is given by  	
\begin{alignat}{1}
\hat{\phi}_2^\star = \begin{cases}
\hat{\phi}_{2,1}   & \text{if } \displaystyle \sum_{n=1}^{N} \mathrm R_{n} \cos(\theta_{n}+ \hat{\phi}_{2,1}) > 0 \\
\hat{\phi}_{2,2}  & \text{otherwise}  \\
\end{cases}, 
\end{alignat}
where $ \theta_{n} = \frac{2(n-1)\pi}{N}$ for $ n \in \mathcal{N} $.	
\end{theorem}
\begin{IEEEproof}
See Appendix \ref{App:Channel Estimation}.
\end{IEEEproof}

Again we should stress that this simplicity in the ambiguity resolving process was made possible due to the methodology we have followed in defining  $\Theta$. The simplicity in our results can be used to further reduce the amount of feedback required to make the proposed scheme work. This reduction will be directly proportional to the resources that you have at the ER. For an example, using the expression in Theorem \ref{esti}, the ER can calculate $\tan{\hat{\phi}_2}$ at the ER, and feedback this value instead of feeding back $N$ RSSI values. Then, the ET can calculate $\hat{\phi}_2$, and request for two further feedback values from the ER to resolve the ambiguity, similar to the method suggested in \cite{rssi_work}. This method effectively reduces the amount of feedback from $N$ to 3. Furthermore, if the ER has enough resources to calculate $\hat{\phi}_2$, as the condition obtained for ambiguity resolution is remarkably simple as well, the ER can directly feedback $\hat{\phi}_2^\star$ to the ET. This method will reduce the amount of feedback from $N$ to 1. These examples give ample evidence to highlight that the results in this paper can be applied and further optimized for many different applications of beamforming. In the next section, we will study the case where $K>2$.

\section{Extension of results for a single-user WET system when $ K > 2 $ } \label{section: N large}

\begin{figure}[t]
	\centering {\includegraphics[scale=0.75]{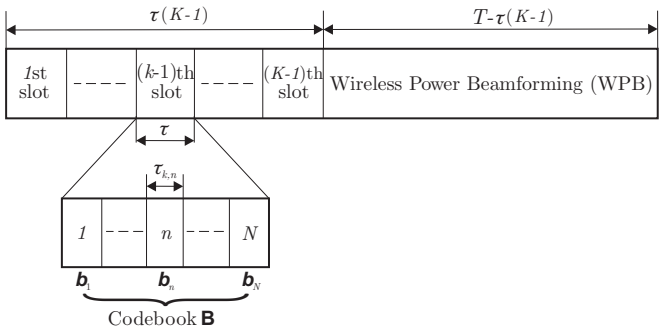}} 
	\caption{The two-phase transmission protocol when $ K>2 $.}	 
	\label{2pha_2}
\end{figure}

When $K>2$, the ET has to estimate $\brparen{\phi_{k}}_{k=2}^{K}$ (refer \eqref{vec}), and for this, we propose a pair wise transmit antenna activation policy. To this end, when a pair of antennas is activated, the phase difference of the channels between the activated antennas and the ER can be estimated by using the same method that we have proposed for $ K=2 $. This pairwise activation is repeated for different antenna pairs until we have estimated  $\brparen{ \phi_{k}}_{k=2}^{K}$. It is not hard to see that the most straightforward way of selecting the best $\brparen{ \phi_{k}}_{k=2}^{K}$ through pairwise activation is by doing an exhaustive search after the activation of all possible antenna pairs, and selecting the WET maximizing $\brparen{ \phi_{k}}_{k=2}^{K}$. However, this approach is too complex to be feasible in practice. Therefore, in this paper, we propose a suboptimal method of estimating $\brparen{ \phi_{k}}_{k=2}^{K}$, that still guarantees satisfactory results. 

The proposed extension is as follows. The training stage is further divided into $ K-1 $ time slots, such that each  time slot consists of $ N $ mini-slots, see Fig. \ref{2pha_2}. This means, there will be $(K-1) \times N$ mini-slots in total in the training stage. When $K=2$, we had only one time slot, and $N$ mini-slots. Let $ \mathcal{K} = \{2,\dots,K\}$. In the $(k-1)$th time slot, where $ k \in \mathcal{K} $, the ET simultaneously activates the $ k $-th antenna and the $ 1 $st antenna, and transmits using each element in $\mathrm {\bf{B}}$. This allows us to estimate $ \phi_k $ using the same method proposed for $ K=2 $ as only a pair of antennas is activated. This also allows us to obtain $\brparen{ \hat \phi_{k}}_{k=2}^{K}$ at the end of the training stage. Then, from \eqref{vec}, the beamforming vector for the WPB stage can be set as $\vec{w}_{\text{WPB}} = \sqrt {\frac{P}{K}} \sqparen {1, e^{-j\hat \phi_{2}} , \dots ,e^{-j\hat \phi_{K}} }^\top$.

It should be noted that the method that we have proposed for estimating $\brparen{\phi_{k}}_{k=2}^{K}$  is a heuristic scheme that activates a pair of antennas at a time. Therefore, by using an example, we will justify the proposed method when compared to the case of jointly estimating $\brparen{\phi_{k}}_{k=2}^{K}$, where the ET simultaneously activates all $ K $ antennas. When $ K=3 $, the RSSI value for the $ n$-th mini slot can be written as
\begin{eqnarray*}
	\mathrm R_n = \alpha_1 + \beta_2 \cos(\theta_n + \phi_2)+ \beta_3 \cos(\theta_n +  \phi_3)+ \\ \beta_{2,3} \cos( \phi_2 - \phi_3) +z_n,
\end{eqnarray*}
where $ \alpha_1 $, $ \beta_2 $, $ \beta_3 $, and $ \beta_{2,3 } $ are the parameters that depend on $ |h_1| $, $ |h_2| $, and $ |h_3| $.
Therefore, the parameter vector becomes $ [\alpha_1 \ \ \beta_2 \ \ \beta_3 \ \ \beta_{2,3} \ \ \phi_2 \ \ \phi_3] $. It can be shown by studying the FIM of the parameter vector that we need at least six RSSI feedback values in order to estimate $ \phi_2 $ and $ \phi_3 $. Even with our proposed pairwise antenna activation policy, we need at least six RSSI feedback values when $ K=3 $. Therefore, we may not achieve a significant feedback reduction. 

Having discussed on the amount of feedback, the greater concern is with the ambiguity resolution. When it comes to phase estimation, ambiguity resolving is a serious practical difficulty. However, we have given a very simple ambiguity resolution procedure in our proposed scheme, without requesting further feedback from the receivers. The ambiguity resolution when $\brparen{\phi_{k}}_{k=2}^{K}$ is estimated jointly, is not at all straightforward. Therefore, the channel learning methodology proposed for $ K>2 $ is still reasonable and justifiable.

\section{The Selection of $ N $} \label{opt_N}

According to the CRLB analysis in Section \ref{section:crlb}, 
we can expect the minimum variance of $ \hat \phi_k $ to scale like  $\frac{2 \sigma^{2}}{N \beta^{2}}$ with $N$. This means, larger $ N $ values yield a higher channel estimation precision. However, larger $ N $ will increase the time spent in training, which will eventually reduce the time for WPB. This may lead to a reduction in the total transferred energy. Therefore, it is not hard to see that $ N $ affects the system performance greatly, and we will focus on setting this important parameter in this section.

We will first derive an expression to approximate the received signal strength in the WPB stage, which we denote as $ \mathrm R_{\mathrm{WPB}} $. When $ K (>2) $, from \eqref{power}, we have
\begin{eqnarray} 
\mathrm R_{\mathrm{WPB}} 
=\alpha_1 +  \displaystyle \sum_{i=2}^{K} \beta_{i} \cos\paren{\Delta \hat \phi_{i}} + \displaystyle \sum_{i=2}^{K-1} \sum_{j=i+1}^{K} \beta_{i,j} \nonumber\\ \cos\paren{\Delta \hat \phi_{i} - \Delta \hat \phi_{j}}, \label{bfore_appor}
\end{eqnarray} 
where $ \Delta \hat \phi_{i} $ and $ \Delta \hat \phi_{j} $ denote the error in estimating $ \phi_{i} $ and $ \phi_{j} $, respectively, for $ i,j \in \{2,\dots,K\} $ and $i \neq j$. 
$\alpha_1 $, $ \beta_{i}  $, and $ \beta_{i,j} $ are the parameters that depend on channel magnitudes between the ET and the ER similar to the ones defined in Section \ref{section: N large}. 
We assume the estimation errors to be small and approximately equal to each other in a given transmission block, \textit{i.e.},  $ \Delta \hat \phi_{i} \approx \Delta \hat \phi_{j} \approx \Delta \hat \phi $. Hence, we have
\begin{alignat} {2}
	\mathrm R_{\mathrm{WPB}} 
	&=\alpha_1 +  \displaystyle \cos\paren{\Delta \hat \phi} \sum_{i=2}^{K} \beta_{i}  + \displaystyle \sum_{i=2}^{K-1} \sum_{j=i+1}^{K} \beta_{i,j}. \label{r_wpb}
\end{alignat}
Since the minimum variance of the estimates behave like  $\frac{2 \sigma^{2}}{N \beta^{2}}$, we can write $\Delta\hat \phi  =  \varepsilon/ \sqrt N  $, where $ \varepsilon $ is a constant. Also, by using the small-angle approximation $ \cos\paren{\Delta \hat \phi} = \paren{1- \frac{\Delta\hat \phi^2}{2}}  $, 
$ \mathrm R_{\mathrm{WPB}} 
= \omega_1 \paren{1 -\frac{\omega_2}{N}}$, 
where $ \omega_1 =  \alpha_1 + \sum_{i=2}^{K} \beta_{i} + \displaystyle \sum_{i=2}^{K-1} \sum_{j=i+1}^{K} \beta_{i,j} $ and $ \omega_2 = \frac{ \paren{\sum_{i=2}^{K} \beta_{i}} \varepsilon^2}{2\omega_1} $. When $ K=2 $, $ \omega_1 =  \alpha_1 + \beta_{2}  $, and when $N \rightarrow \infty$, $\mathrm R_{\mathrm{WPB}}$ turns out to be $\omega_1$. 


\begin{figure}[t] \vspace{0.15cm}
	{\includegraphics[trim = 5mm 0mm 0mm 10mm, clip, scale=0.5]{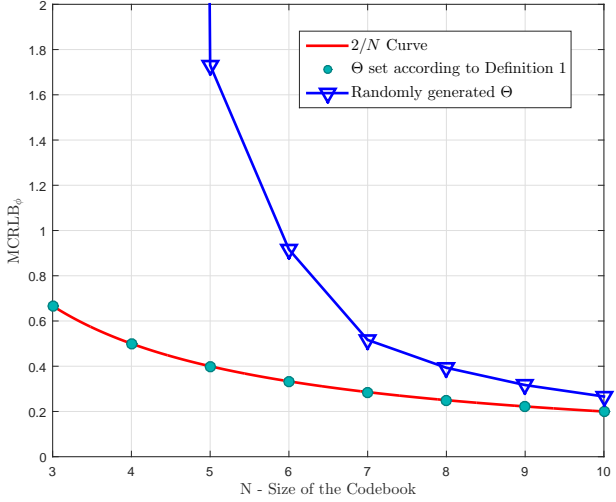}} 
	\caption{The behavior of the $ \mathrm{MCRLB}_{\phi}$ with N when $\beta=\sigma=1$.}	 
	\label{mcr}
\end{figure}

Let $ E_f $ and $ \tau_{k,n} $ be the energy required and time required to feed back a single RSSI value, respectively. Hence, $ N(K-1)\tau_{k,n} $ is the time taken for the training stage. If the transmission block length is $ T $, the harvested energy during a single transmission block can be written as
\begin{alignat} {2}
	\mathrm E_{\mathrm{total}} 
	&= \Big(T- N(K-1)\tau_{k,n}\Big)  \mathrm R_{\mathrm{WPB}} -N(K-1)E_f \nonumber \\
	&= \omega_1  \Big(T- N(K-1)\tau_{k,n}\Big)   \paren{1 -\frac{\omega_2}{N}} -N(K-1)E_f. \label{E_total}
\end{alignat}
Using this expression, we will provide bounds for the optimal value of $N$ through the following theorem.  
\begin{theorem}\label{N_range}
Let $ N^\star $ be the optimal value of $N$, and $ T > N(K-1)\tau_{k,n} $. Then,  $ 3 \leq N^{\star} \leq \sqrt{ \frac{3T}{(K-1)}} $.	
\end{theorem}
\begin{IEEEproof}
For  positive values of $ N $, \eqref{E_total} is convex. By differentiating $  \mathrm E_{\mathrm{total}} $ with respect to $ N $ and setting it to zero, the optimal value of $ N $ that maximizes $ \mathrm E_{\mathrm{total}} $ can be given as 
\begin{alignat} {2}
N^{\star} 
&= \sqrt{\psi \frac{L}{(K-1)}},
\label{opt_n} 
\end{alignat}
where $ \psi = \frac{\omega_1 \omega_2 \tau_{k,n}}{\omega_1 \tau_{k,n} + E_f} $. We need at least 3 RSSI feedback values to estimate the channels between the ET and the ER. Therefore, the lower bound of $ N^{\star} $ is 3. If $ T $ is long enough to harvest energy, $ \mathrm E_{\mathrm{total}}  $ is strictly positive. Therefore, we have
\begin{alignat*} {2}
0 &< \omega_1 \Big(T- 3(K-1)\tau_{k,n}\Big)   \paren{1 -\frac{\omega_2}{3}} -3(K-1)E_f \nonumber\\
\Rightarrow \omega_2 &< 3 - \frac{9 (K-1) E_f}{\omega_1  (T-3(K-1)\tau_{k,n})} < 3, 
\end{alignat*}	
because $ \omega_1$, $ E_f$, $\tau_{k,n} >0 $, and $ T > 3(K-1)\tau_{k,n} $. Also, since  $ \frac{\omega_1  \tau_{k,n}
	}{\omega_1 \tau_{k,n} + E_f} < 1$, $ \psi  <3 $. Therefore, from \eqref{opt_n}, the upper bound of the $ N^{\star} $ is $ \sqrt{ \frac{3T}{(K-1)}} $, which completes the proof. 
\end{IEEEproof}
In the next section, we will validate our results using numerical
evaluations.

\begin{figure}[t] 
	{\includegraphics[trim = 7mm 0mm 0mm 10mm, clip, scale=0.45]{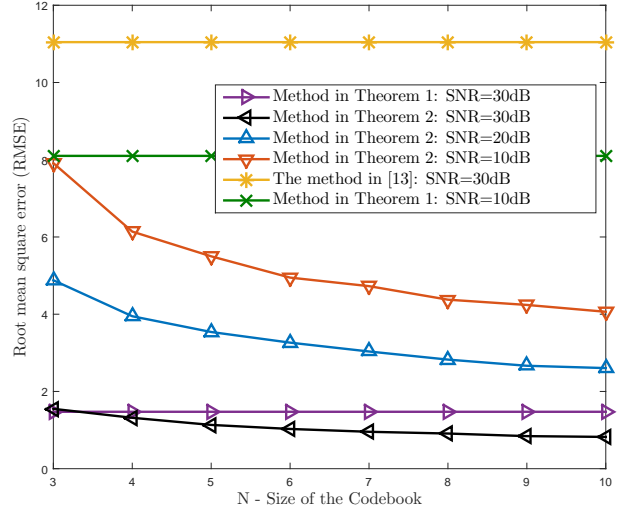}} 
	\caption{The behavior of the root mean squared error (RMSE) of $ \hat \phi_2 $ for
		different SNR values when $\beta=\sigma=1$.}	 
	\label{ML}
\end{figure}

\section{Numerical Evaluations} \label{section:Numerical Evaluations}

In this section, we present some numerical examples to validate our
proposed schemes, and to provide useful insights on channel learning and
wireless power beamforming. As a start, in Lemma \ref{lemma3}, we have
focused on $ \mathrm{MCRLB}_{\phi} $,  and we have given the formal proof
for the minimum $ \mathrm{MCRLB}_{\phi} $ value, considering $ N=3 $ and $ N=4 $,
respectively. Then, based on the pattern, we expected that the minimum $
\mathrm{MCRLB}_{\phi}$ to take the form of $ \frac{2 \sigma^{2}}{N
\beta^{2}} $ for arbitrary values of $ N $. Validation of this result is
presented in Fig. \ref{mcr}. For the numerical evaluations, we have set
$\beta = \sigma =1 $, and we have calculated $\mathrm{MCRLB}_{\phi} $
according to Lemma \ref{lemma2}, while setting the phase values according
to $\Theta$ in Definition 1. We can see that setting the phase values
according to Definition 1 allows us to achieve the minimum MCRLB as the
values lie on the $ 2/N $ curve. The figure also shows how the average $
\mathrm{MCRLB}_{\phi} $ behaves if the phase values in $ \Theta $ are
chosen randomly, for a given $ N $. It can be seen that the average $
\mathrm{MCRLB}_{\phi} $ values lie above the $ 2/N $ curve, with the gap
reducing when $N$ is increased. Due to this reason, a $
\mathrm{MCRLB}_{\phi} $ value obtained by a randomly generated $ \Theta $
can be achieved using a lower number of feedback values, if $ \Theta $ is
defined according to Definition 1. This is vital as we are dealing with a
receiver having a tight energy constraint, and we have to also minimize the
time spent for the training stage. Finally, as expected, we can observe
that when $ N $ increases, the lower bound on the variance of $
\hat{\phi_2} $ decreases.

\begin{figure}[t] 
	{\includegraphics[trim = 8mm 0mm 0mm 8mm, clip, scale=0.545]{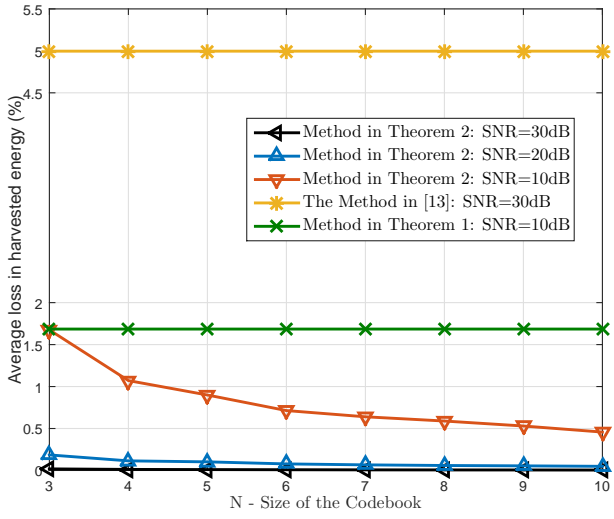}} 
	\caption{The behavior of the average percentage loss in harvested energy for
		different SNR values when $\beta=\sigma=1$.}	 
	\label{av_energy}
\end{figure}

\begin{figure}[t] 
	{\includegraphics[trim = -5mm 0mm 0mm 0mm, clip, scale=0.45]{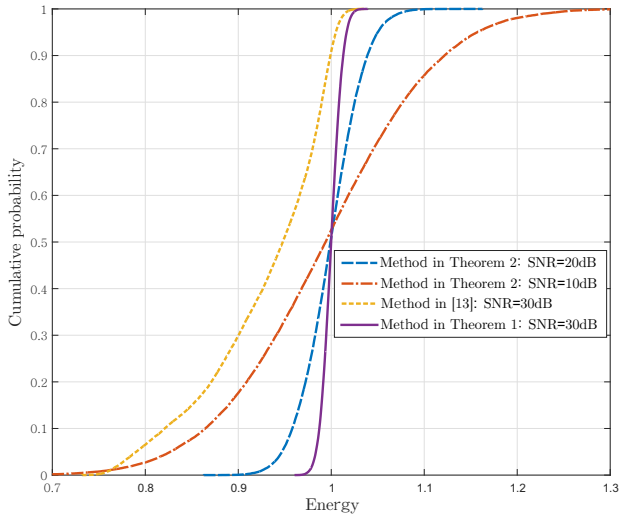}} 
	\caption{Empirical CDF illustrations of energy  when $ K=2 $.}	 
	\label{energy4}
\end{figure}

\begin{figure}[t] \vspace{-0.4cm}
	{\includegraphics[trim = 8mm 0mm 0mm 0mm, clip, scale=0.545]{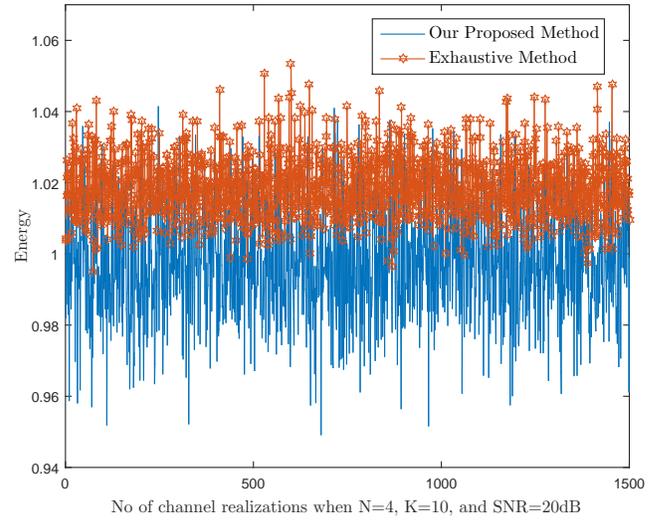}} 
	\caption{Comparison between proposed and baseline method for single user channel learning and WPT when $ K>2 $.}	 
	\label{ehhau}
\end{figure}

\begin{figure}[t]
	{\includegraphics[trim = 8mm 7mm 0mm 11mm, clip, scale=0.485]{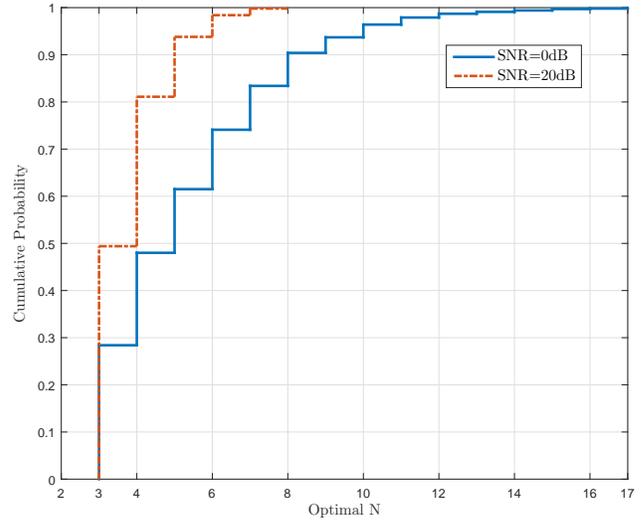}} 
	\caption{The behaviour of  $ N^\star $, when $ K=2 $ and $ T $ is $ 100 $ times longer than $ \tau_{k,n} $.}	 
	\label{opt_NN} \vspace{-0.5cm}
\end{figure}

In Theorem \ref{esti}, we have presented an ML estimate of $ \phi_2 $.
Fig. \ref{ML} illustrates the behavior of the root mean squared error
(RMSE) with $ N $ for different SNR values. $\Theta$ is defined according
to Definition 1. As expected, for higher SNR values, we have lower RMSE
values, and the RMSE values converge to zero with $N$. It is interesting to
note that even when $ N=3 $, the phase error is not significantly large.
For example, when $ N=3 $ and SNR$ =10 $dB, RMSE is $ 7.88^{\circ}$. It is
also interesting to note that when $ N=3 $, the RMSE of calculating phase
values according to Theorem \ref{esti} is approximately equal to the RMSE
of calculating phase values according to the  method
proposed in Theorem \ref{th_no_noise}, where the effect of noise is neglected. Furthermore, Fig. \ref{ML}
illustrates that our proposed method allows the ET to achieve significant
gains when compared to other works in the literature, even with lower SNR
values. 

Fig. \ref{av_energy} illustrates the average loss in harvested energy (percentage) due to using the the proposed methodology, compared to performing  energy beamforming  with perfect CSI. We can see that the loss is rather acceptable given the practicality of the proposed method. Fig. \ref{energy4} illustrates the respective energy transfer performance of each case considered in Fig. \ref{ML}, using empirical
cumulative distribution functions. This alternative form of representation is
used for improved clarity. The important point to notice in the figure is that the
variance has decreased with SNR. This is because the increase in SNR leads to a better estimation, and the ET can guarantee a certain energy transfer with a high probability, that is, lower outage.

In Section \ref{section: N large}, we have extended the proposed channel learning and WPB scheme for $ K>2 $ using a suboptimal, but energy efficient method. A simple test was carried out in order to check the performance of the proposed method. When $ \xi=1 $, $ N=4 $, $ K=10 $ and SNR$=20  $dB, we randomly generated $\brparen{\theta_n}_{n=1}^4$ and $\brparen{\phi_k}_{k=2}^{10}$ assuming that they are uniformly distributed between 0 and 2$\pi$, and $ \vec{w}_{\text{WPB}} $ is calculated using the exhaustive method and the proposed method, respectively. As shown in Fig. \ref{ehhau}, this was done for 1500 channel realizations. Although the feedback load is reduced considerably, it is not hard to see that our proposed method still exhibits impressive results. On average, the loss is only 2.2$\%$. However, we can see that the variance has increased by shifting to the suboptimal method, similar to what was highlighted using Fig. \ref{energy4}. 

We should note that although larger $ N $ yields a higher channel estimation precision, this reduces the time for WPB. Therefore, a larger $N$ may lead to a reduction in the total transferred energy. In Section \ref{opt_N}, we have obtained  bounds for the  value of $N$ that maximizes the energy transfer during the WPB stage, \textit{i.e.} bounds on $ N^\star $. Fig. \ref{opt_NN} illustrates the behaviour of the CDF of $ N^\star $, when $ K=2 $ and $ T=100 \tau_{k,n} $. For these parameters, from Theorem \ref{N_range}, the theoretical lower bound and the upper bound are 3, and 17, respectively. Fig. \ref{opt_NN} is consistent with these results and depict that the bounds are tight as well. We can observe that when the SNR increases, the CDF shifts to the left. This is because for better channels, we need less feedback for an accurate estimation, and hence, the optimal $N$ will lie closer to the lower bound of the region with a   higher probability. Having done the numerical evaluations, we will further validate our results experimentally in the next section.

\section{Experimental Validation}\label{Experimental Validation}

%

\begin{figure}[t] 
	\centering{\includegraphics[scale=0.5]{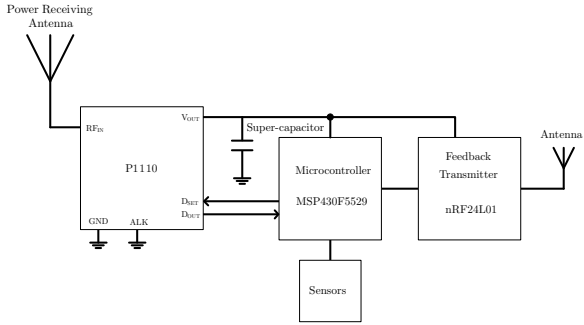}} 
	\caption{The hardware block diagram of the ER.}	 
	\label{Fig: ER} 
\end{figure}

\begin{figure}[t]
	\centering {\includegraphics[scale=0.462]{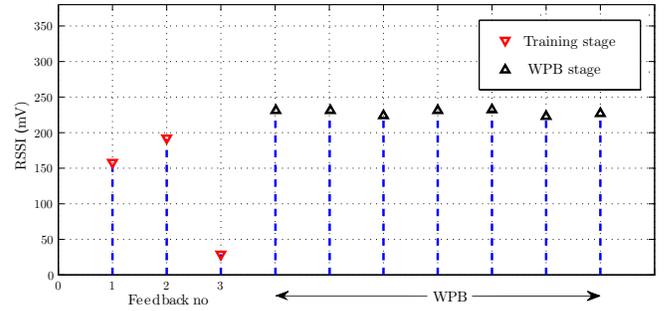}} 
	\caption{The RSSI values corresponding to each stage when the ET and the ER are 2m apart and $ N=3 $.}	 
	\label{Fig: rssi_3} 
\end{figure}

In our experimental setup, the ET consists of $2$ antennas and delivers energy to an ER consisting of a single antenna. The implementation of our ER is shown in Fig. \ref{Fig: ER}. 
We use Powercast P1110 power-harvester, 
which has an operating band ranging from 902 to 928MHz. P1110 has an analog output ($ \mathrm {D_{OUT}} $), which provides an analog voltage level corresponding to the RSSI. 
As the storage device of our design, we use a low leakage
0.22F super-capacitor. 
The output of P1110 charges the super-capacitor and the super-capacitor powers the microcontroller, the feedback transmitter and the sensors. 
An Ultra-Low-Power MSP430F5529 microcontroller is used to read the RSSI values and transmit them via the feedback transmitter. 
When functioning, the microcontroller and the feedback transmitter are on sleep mode, and after each 500 ms interval, both wake up from sleep in order to read the RSSI and transmit it to the ET. NORDIC nRF24L01 single chip 2.4GHz transceiver has been used as the feedback transmitter. 
When the ER operates in active mode (reading RSSI values and transmitting), it consumes only 12.8 $ \mu $J/ms and it consumes negligible energy in sleep mode. 
The SDR used in our ET is USRP B210, which has $2  \times  2$ MIMO capability. CRYSTEC RF power amplifiers (CRBAMP 100-6000) are used to amplify the RF power output of the USRP B210. 
All the real-time signal processing tasks, channel phase difference ($ \phi_k $) estimation and setting beamforming vectors in both training and WPB stages were performed on a laptop using the GNU Radio framework. 
We use 915Mhz as the beamforming frequency. The same transceiver chip used in the ER, nRF24L01, is used as the feedback receiver at the ET side. For the experiment, the ET and the ER are 2 meters apart. Using this setup, for $ N=3 $, Fig. \ref{Fig: rssi_3} illustrates the training stage and the WPB stage, and we can see a clear gain by the proposed method.  

\begin{figure}[t]
	\centering {\includegraphics[scale=0.47]{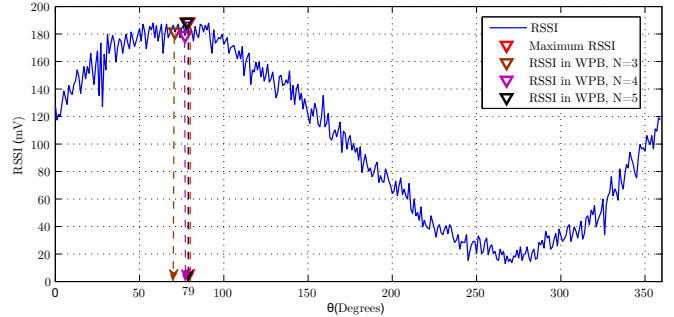}} 
	\caption{The RSSI values when $ \theta_{i} $ is changed from $ 0^{\circ} $ to $ 360^{\circ} $  with $ 1^{\circ} $ resolution.}	 
	\label{Fig: 0-360} 
\end{figure} 

\begin{table} 
	\caption {Experimental results} 
	\rowcolors{1}{}{lightgray}
	\centering \begin{tabular}{  p{1.5cm}  p{1.5cm}  p{1.7cm}  }
		\hline \centering  N & \centering $ \hat \phi_2 $ & Error $ | \hat \theta -79^{\circ}| $ \\
		\hline \centering  3 & \centering $ 71^{\circ} $ & $ \qquad 8^{\circ} $ \\
		\centering  4 & \centering $ 77^{\circ} $  & $\qquad 2^{\circ} $ \\
		\centering  5 & \centering $ 78^{\circ} $ & $\qquad 1^{\circ} $ \\ 
		                        	
	\end{tabular} 
	
\end{table} 

Then, we focused on validating the result on phase estimation. For this, we changed $ \theta_{n} $ from $ 0 $ to $ 360 $ degrees with $ 1^{\circ} $ resolution, and collected all respective RSSI values (see Fig. \ref{Fig: 0-360}). Since it was not practical to collect all the 360 RSSI values using the harvested energy via the feedback transmitter, we used a wired feedback for this experiment. 
Fig. \ref{Fig: 0-360} shows that the maximum RSSI occurs when $ \theta_{n} = 79^{\circ}$. Therefore, the maximum energy transfer happens at that point. 
Using the same set of values, we estimated $\hat \phi_2 $ ($\Theta$ defined according to Definition 1) for $ N=3 $, $ N=4 $, $ N=5 $ and $ N=6 $, respectively. The results are tabulated in Table I. It is not hard to see that the errors are significantly small, and they are consistent with the numerical evaluations as well. Further, by using our proposed scheme, and based on the assumption that the conversion efficiency of the power-harvester is fixed, we can extend the range of the ER by 52\% on average. This has been calculated based on the experimental results considering free space loss.  

\section{Conclusions}\label{Section:Conclusions}

This paper has proposed a novel channel estimation methodology to be used in a multiple antenna single user WET system. The ET transmits using beamforming vectors from a codebook, which has been pre-defined using a Cramer-Rao lower bound analysis. RSSI value corresponding to each beamforming vector is fed back to the ET, and these values have been used to estimate the channel through a maximum likelihood analysis. The channel estimation has then been used to set the beamforming vector for the WET. The results that have been obtained are simple, requires minimal processing, and can be easily implemented. The paper has also studied how the estimation ambiguities can be resolved in an energy efficient manner. The analytical results in the paper have been validated numerically, as well as experimentally, while providing interesting insights. It has been shown that the results in the paper are more appealing compared to existing multiple antenna channel estimation methods in WET, especially when there are tight energy constraints and hardware limitations at the ER. Also, the methods can be used for many applications of beamforming, where processing capabilities of the receiver is limited. 

\appendices
\section{Cramer-Rao Lower Bound Analysis } \label{App:crlb}

\subsection{Worst-case CRLB Performance} 
\begin{lemma}
The Gaussian distribution minimizes/maximizes the FIM/CRLB of $\vec \varphi  $.
\end{lemma}
\begin{IEEEproof}
The log likelihood function of \eqref{final_model} can be written as
\begin{eqnarray*}
	l(\mathrm{\bf{R}} , \vec x_{\varphi} ) = \log {f_{\mathrm{\bf{R}} | \vec x_{\varphi} }} (\mathrm{\bf{R}} , \vec x_{\varphi}),
\end{eqnarray*} 
where $ {f_{\mathrm{\bf{R}} | \vec x_{\varphi} }} (\mathrm{\bf{R}} , \vec x_{\varphi}) $ denotes the conditional density function of $ \mathrm{\bf{R}} $ given $ \vec x_{\varphi}  $. Since $\vec x_{\varphi} $ and $\vec z $ are two independent vectors, $ {f_{\mathrm{\bf{R}} | \vec x_{\varphi} }} (\mathrm{\bf{R}} , \vec x_{\varphi}) = f_{\vec z}(\mathrm{\bf{R}} - \vec x_{\varphi})$, where $ f_{\vec z}(\cdot) $ denotes the density function of $ \vec z $.
Now, the first derivative of the log likelihood function can be written as
\begin{alignat*}{4}
\frac{\partial l(\mathrm{\bf{R}} , \vec x_{\varphi} )}{\partial \varphi}  &= \frac{\partial l(\mathrm{\bf{R}} - \vec x_{\varphi} ) }{\partial \varphi} \nonumber\\
&= -\frac{\partial \vec x_{ \varphi} }{\partial \varphi} \frac{\partial l(\mathrm{\bf{R}} - \vec x_{\varphi} )}{\partial \vec z} \nonumber\\
&= -\frac{\partial \vec x_{ \varphi} }{\partial \varphi} \frac{\partial l(\vec z )}{\partial \vec z}. 
\end{alignat*}
${\mathrm {FIM}}_{\varphi}(\mathrm{\bf{R}}) $ is defined as the covariance matrix of $\displaystyle \frac{\partial l(\mathrm{\bf{R}}, \vec x_{\varphi} ) }{\partial \varphi}  $, \textit{i.e.}, 
\begin{alignat}{3}
{\mathrm {FIM}}_{\varphi}(\mathrm{\bf{R}}) & =  \mathbb{E}_{x_{\varphi},z} \bigg[ \bigg(\displaystyle \frac{\partial l(\mathrm{\bf{R}} , \vec x_{\varphi} ) }{\partial \varphi}\bigg) \bigg(\displaystyle \frac{\partial l(\mathrm{\bf{R}} , \vec x_{\varphi} )}{\partial \varphi}\bigg)^{T} \bigg] \nonumber \\
& = \mathbb{E}_{x_{\varphi},z} \bigg[\frac{\partial \vec x_{ \varphi} }{\partial \varphi}\paren {\frac{\partial l(\vec z ) }{\partial \vec z} \frac{\partial l(\vec z )^{T}}{\partial \vec z}} \frac{\partial \vec x_{ \varphi}^{T}}{\partial \varphi} \bigg]. \nonumber
\end{alignat} 
Since $\vec x_{\varphi}  $ and $ \vec z $ are two independent vectors, 
\begin{alignat}{3}
{\mathrm {FIM}}_{\varphi}(\mathrm{\bf{R}}) & = \mathbb{E}_{x_{\varphi}} \bigg[\frac{\partial \vec x_{ \varphi} }{\partial \varphi} \mathbb{E}_{z} \bigg [\frac{\partial l(\vec z ) }{\partial \vec z} \frac{\partial l(\vec z )^{T}}{\partial \vec z} \bigg] \frac{\partial \vec x_{ \varphi}^{T}}{\partial \varphi} \bigg] \nonumber \\
& = \mathbb{E}_{x_{\varphi}} \bigg[\frac{\partial \vec x_{ \varphi} }{\partial \varphi}  \mathrm {FIM}(\vec z) \frac{\partial \vec x_{ \varphi}^{T}}{\partial \varphi} \bigg], \nonumber
\end{alignat} 
where $\mathrm {FIM}(\vec z) $ is the FIM with respect to $\vec z$. Let $ \tilde{\vec z} $ denote a non-Gaussian vector having same size as $ \vec z $. We have $ \mathrm {FIM}(\tilde{\vec z}) \geqslant \mathrm {FIM}(\vec z) $ \cite{fisher}. This implies that 	
\begin{alignat}{2} 
\mathbb{E}_{x_{\varphi}} \bigg[\frac{\partial \vec x_{ \varphi}}{\partial \varphi}  \mathrm {FIM}(\tilde{\vec z}) \frac{\partial \vec x_{ \varphi}^{T}}{\partial \varphi} \bigg] & \geqslant \mathbb{E}_{x_{\varphi}} \bigg[\frac{\partial \vec x_{ \varphi}}{\partial \varphi}  \mathrm {FIM}({\vec z}) \frac{\partial \vec x_{ \varphi}^{T}}{\partial \varphi} \bigg]. \nonumber
\end{alignat}
Therefore, the Gaussian distribution minimizes FIM of $\vec \varphi  $. Also, the Gaussian distribution maximizes the CRLB of $\vec \varphi  $ since the CRLB is given by the inverse of the FIM, which completes the proof.
\end{IEEEproof}

\subsection{Proof of Lemma \ref{Lemma: FIM}}
We have
	\begin{equation}
	\frac{\partial \vec x_{\varphi} }{\partial \vec \varphi}
	= 
	\begin{bmatrix}
	1 & A_1  & D_1 \\
	\vdots & \vdots & \vdots \\
	1 & A_N  & D_N
	\end{bmatrix}.
	\label{mat_j}
	\end{equation}
	By using the FIM of a Gaussian random vector in \cite{crlb_book}, and using the fact that $\mathrm {\bf{C} }_{zz}$ is independent of $\vec \varphi$, the FIM of $\mathrm{\bf{R}}$ can be written as
	\begin{equation*} 
	\mathrm{FIM}_{\varphi}(\mathrm{\bf{R}}) = \bigg[ \frac{\partial \vec x_{\varphi} }{\partial \vec \varphi}   \bigg]^{\top} \mathrm {\bf{C} }_{zz} \bigg[ \frac{\partial \vec x_{\varphi} }{\partial \vec \varphi}   \bigg] . 
	\end{equation*}
	Substituting from \eqref{mat_j} completes the proof. 	 

\subsection{Proof of Lemma \ref{Lemma: CRLB of phi}}
When  $ N\geq 3 $ and $\Theta$ has $ N $ distinct elements, $ \det(\mathrm{FIM}_{\varphi}(\mathrm{\bf{R}})) \neq 0$, and $ \mathrm{FIM}_{\varphi}(\mathrm{\bf{R}}) $ is invertible. Therefore, computing the third diagonal element of the inverse of $\mathrm{FIM}_{\varphi}(\mathrm{\bf{R}})$ completes the proof.

\subsection{Proof of Lemma \ref{lemma3}}

By differentiating \eqref{crlb2}  with respect to $ \theta_{2} $ and $ \theta_{3} $, respectively, and by setting $\theta_{1}=0$, we obtain two expressions which are functions of $ \theta_{2}$ and $ \theta_{3}$. Equating the two expressions to zero and simultaneously solving them under the constraints $\theta_{2},\theta_{3} \in (0,2\pi]$, and  $\theta_{1} \neq \theta_{2} \neq \theta_{3} $, gives us $ \theta_{2}= 2\pi /3 $ and $ \theta_{3}= 4\pi/3 $. Evaluating the Hessian matrix at the stationary point $ \paren{0 , 2\pi /3, 4\pi/3 } $ shows that the stationary point is a minimum. Substituting $ \paren{0 , 2\pi /3, 4\pi/3 } $ in \eqref{crlb2} gives us $2\sigma^{2}/3 \beta^{2}$, which completes the proof for $N=3$. Following the same lines for $N=4$ completes the proof of the lemma.

\section{Estimation of $ \phi_2 $} \label{App:Channel Estimation}

\subsection{Proof of Lemma \ref{aux}}

Let    $ \displaystyle \theta_{n} = (n-1)\tilde{\theta}_N $, where $ \tilde{\theta}_N =\displaystyle\frac{2\pi}{N} $.	Now,
\begin{equation*}
\begin{array}{l@{}l}
\displaystyle \sum_{n=1}^{N} \sin(\theta_{n}+\phi_2) & =\displaystyle \left. \sin(\phi_2)+\dots+\sin((N-1)\tilde{\theta}_N+\phi_2) \right. \\
& = \displaystyle \operatorname{Im} \Big\{\paren{ 1 +\dots+ e^{j((N-1)\tilde{\theta}_N)}}e^{j\phi_2}  \Big\} \\
& =\displaystyle \operatorname{Im} \Big\{ \frac{(1-e^{jN\tilde{\theta}_N})e^{j\phi_2}}{\displaystyle 1-e^{j\tilde{\theta}_N}}  \Big\} .
\end{array}
\end{equation*}
We have $e^{jN\tilde{\theta}_N}= e^{j 2 \pi}=1$. Hence, $\sum_{n=1}^{N} \sin(\theta_{n}+\phi_2)=0$.
Following similar steps for the other summations of interest  completes the proof.

\subsection{Proof of Theorem \ref{esti}}
When $ \theta_{n} = {2(n-1)\pi}/{N}$, from Lemma \ref{aux}, we have $\sum_{n=1}^{N} \sin(\theta_{n}+\phi_2)=\sum_{n=1}^{N} \sin{[2(\theta_{n}+\phi_2)]}=0 $.  Therefore, \eqref{Estimation_phi} can be simplified and written as $ \sum_{n=1}^{N} \mathrm R_{n}\sin{(\theta_{n}+\phi_2)} = 0  $, which is independent of $\alpha$ and $\beta$. By expanding $\sin{(\theta_{n}+\phi_2)}$ we get,
\begin{equation*}
\sin \phi_2 \sum_{n=1}^{N} \mathrm R_{n}\cos{\theta_{n}} + \cos \phi_2 \sum_{n=1}^{N} \mathrm R_{n}\sin{\theta_{n}} = 0 ,
\end{equation*}
which can be directly used to obtain \eqref{estimation_phi}, completing the proof.

\subsection{Proof of Theorem \ref{ambi}}
	By taking the second derivative of \eqref{ml} with respect to $ \phi_2 $, we have 
	\begin{multline*}
	\frac{\partial^2 \mathrm{E}}{\partial \phi^2_2}	= \sum_{n=1}^{N} \mathrm R_{n}\cos{(\theta_{n}+\phi_2)} - \alpha \sum_{n=1}^{N} \cos{(\theta_{n}+\phi_2)} \\ - {\beta} \sum_{n=1}^{N} \cos{[2(\theta_{n}+\phi_2)]}.	\label{2diff}	
	\end{multline*}
	When $ \theta_{n} = {2(n-1)\pi}/{N}$, from the Lemma \ref{aux}, we have $\sum_{n=1}^{N} \cos(\theta_{n}+\phi_2)=\sum_{n=1}^{N} \cos{[2(\theta_{n}+\phi_2)]}=0 $. Therefore, 
	\begin{dmath}
		 \frac{\partial^2 \mathrm{E}}{\partial \phi^2_2}	= \sum_{n=1}^{N} \mathrm R_{n}\cos{(\theta_{n}+\phi_2)}, 	\label{2ndderi}
	\end{dmath}	
	 which is again independent of $\alpha$ and $\beta$. Now, 
	 \begin{dmath*}
	 \brparen{	\frac{\partial^2 \mathrm{E}}{\partial \phi^2_2} }_{\phi_2= \hat{\phi}_{2,1}}	= \sum_{n=1}^{N} \mathrm R_{n}\cos{(\theta_{n}+\hat{\phi}_{2,1})}, 	
	 \end{dmath*}	
	 and
	 \begin{dmath*}
	 	\brparen{	\frac{\partial^2 \mathrm{E}}{\partial \phi^2_2} }_{\phi_2= \hat{\phi}_{2,2} = (\hat{\phi}_{2,1}-\pi)}	= -\sum_{n=1}^{N} \mathrm R_{n}\cos{(\theta_{n}+\hat{\phi}_{2,1})}. 	
	 \end{dmath*}	
	 It is not hard to see that if \eqref{2ndderi} is positive for one possible solution, then \eqref{2ndderi} is negative for the other possible solution. Moreover, as discussed in Theorem \ref{esti}, $\hat{\phi}_{2,1}$  and $\hat{\phi}_{2,2}$ are critical points of \eqref{ml} (first derivative of \eqref{ml} was zero at these points). Therefore, we can claim that one of the candidate solutions is a local minima, while the other is a local maxima. Since we want to find the local minima of $E$ which maximizes the RSSI, the solution which satisfies the second derivative test for the local minima gives us the correct estimate $ \phi_2^\star $, which completes the proof.

\bibliography{bibfile}

	 
	 \begin{IEEEbiography}[{\includegraphics[width=1in,height=1.25in,clip,keepaspectratio]{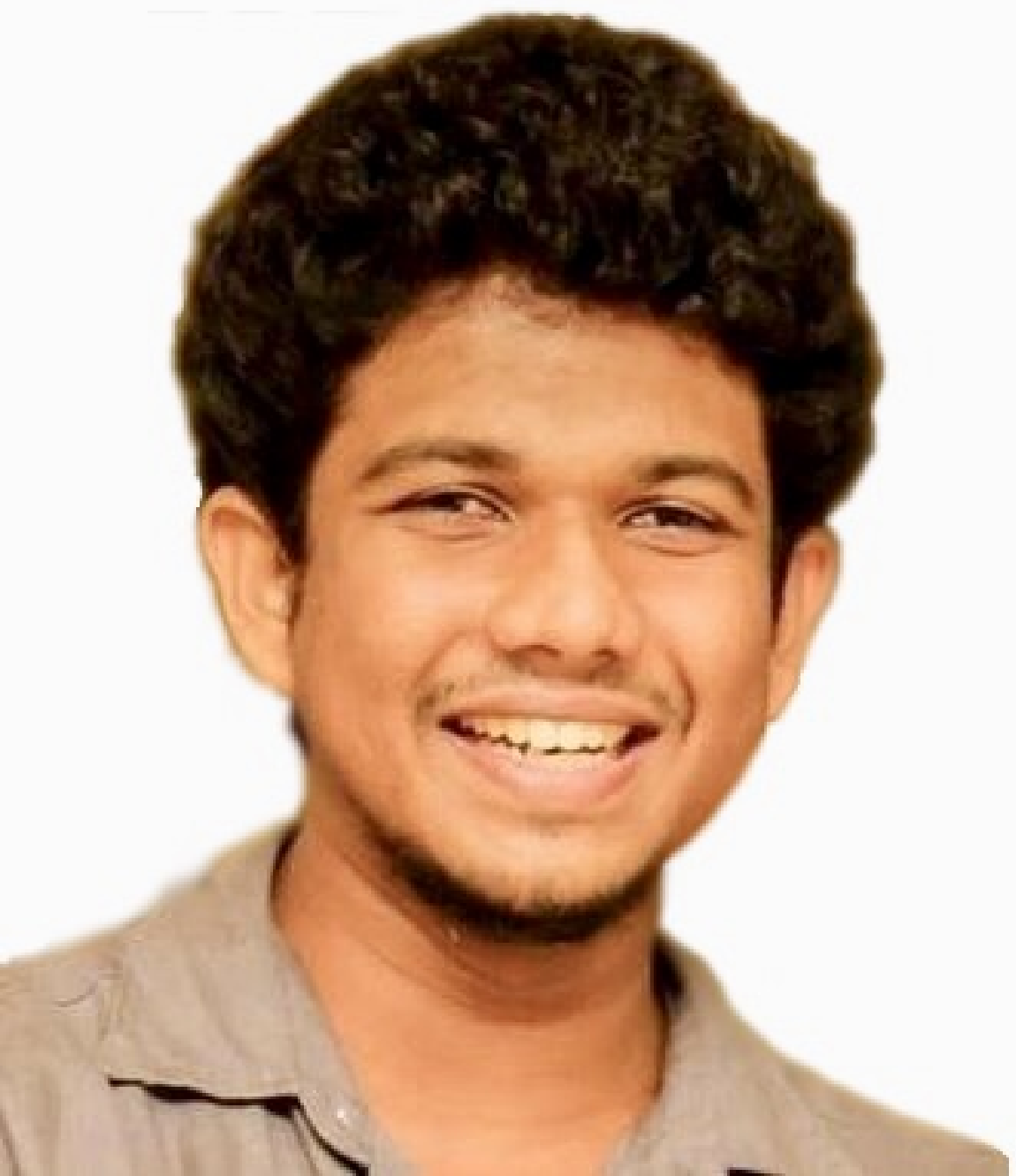}}]{Samith Abeywickrama}
	 	 (S'16) received the B. Sc. degree in engineering from the Department of Electronic and Telecommunication Engineering, University of Moratuwa, Sri Lanka, in 2016. He then joined with Singapore University of Technology and Design, as a researcher. His research interests include proof-of-concept and taking advanced theoretical ideas all the way to practice using software-defined radios (SDR), wireless power transfer, wireless spectrum sensing, and angle of arrival (AOA) estimation.
	 \end{IEEEbiography}
	 
	 \begin{IEEEbiography}[{\includegraphics[width=1in,height=1.25in,clip,keepaspectratio]{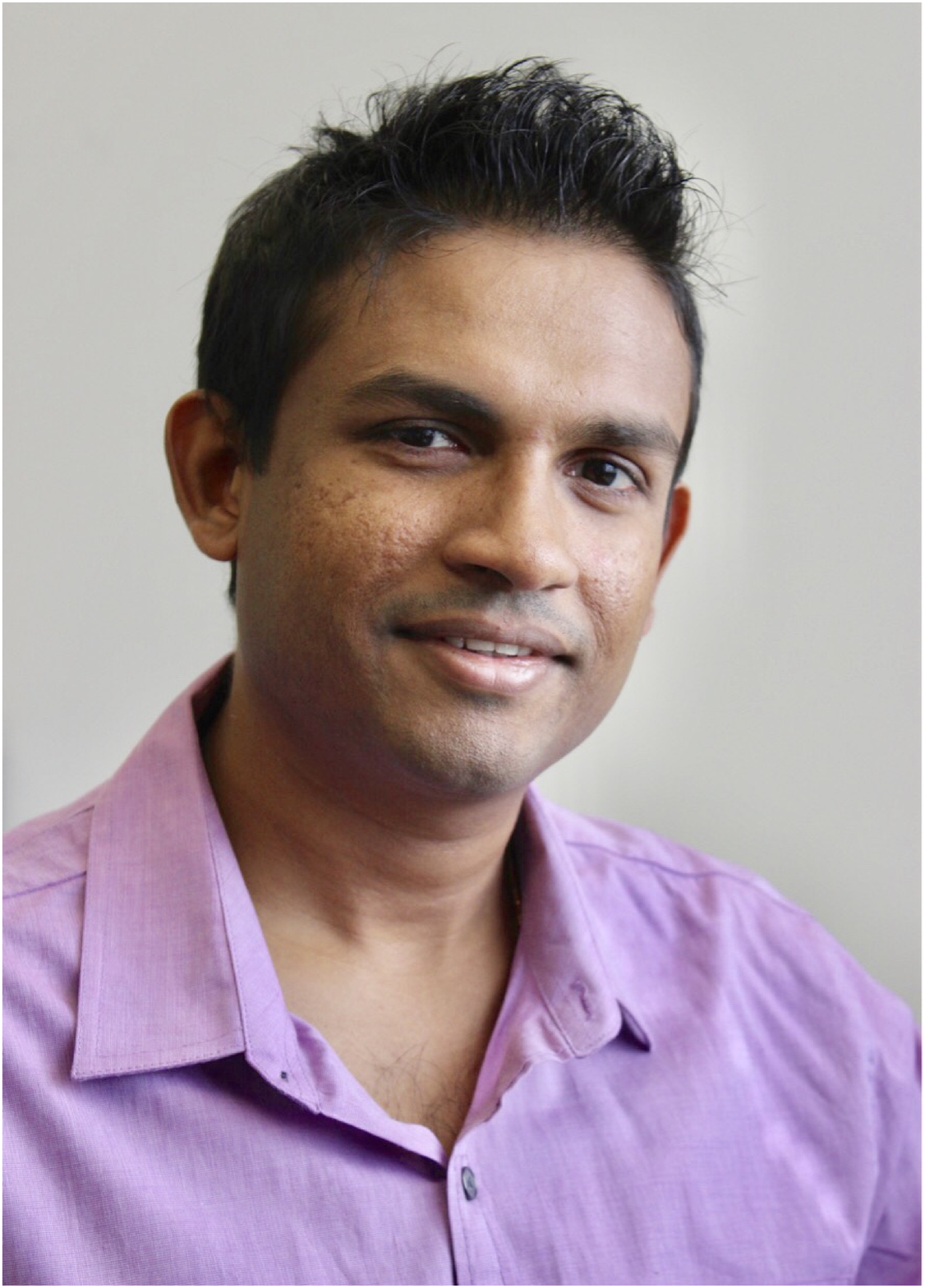}}]{Tharaka Samarasinghe }
	 	(S'11-M'13) was born in Colombo, Sri Lanka, in 1984. He received the B.Sc. degree in engineering from the Department of Electronic and Telecommunication Engineering, University of Moratuwa, Sri Lanka, in 2008, where he received the award for the most outstanding undergraduate upon graduation. He received the Ph.D. degree from the Department of Electrical and Electronic Engineering, University of Melbourne, Australia, in 2012. He was a Research Fellow at the Department of Electrical and Computer Systems Engineering, Monash University, Australia, from 2012 to 2014. He has been with the Department of Electronic and Telecommunication Engineering, University of Moratuwa, Sri Lanka, since January 2015, where he is currently a Senior Lecturer. His research interests are in communications theory, information theory, and wireless networks.
	 \end{IEEEbiography}
	 
	 \begin{IEEEbiography}[{\includegraphics[width=1in,height=1.25in,clip,keepaspectratio]{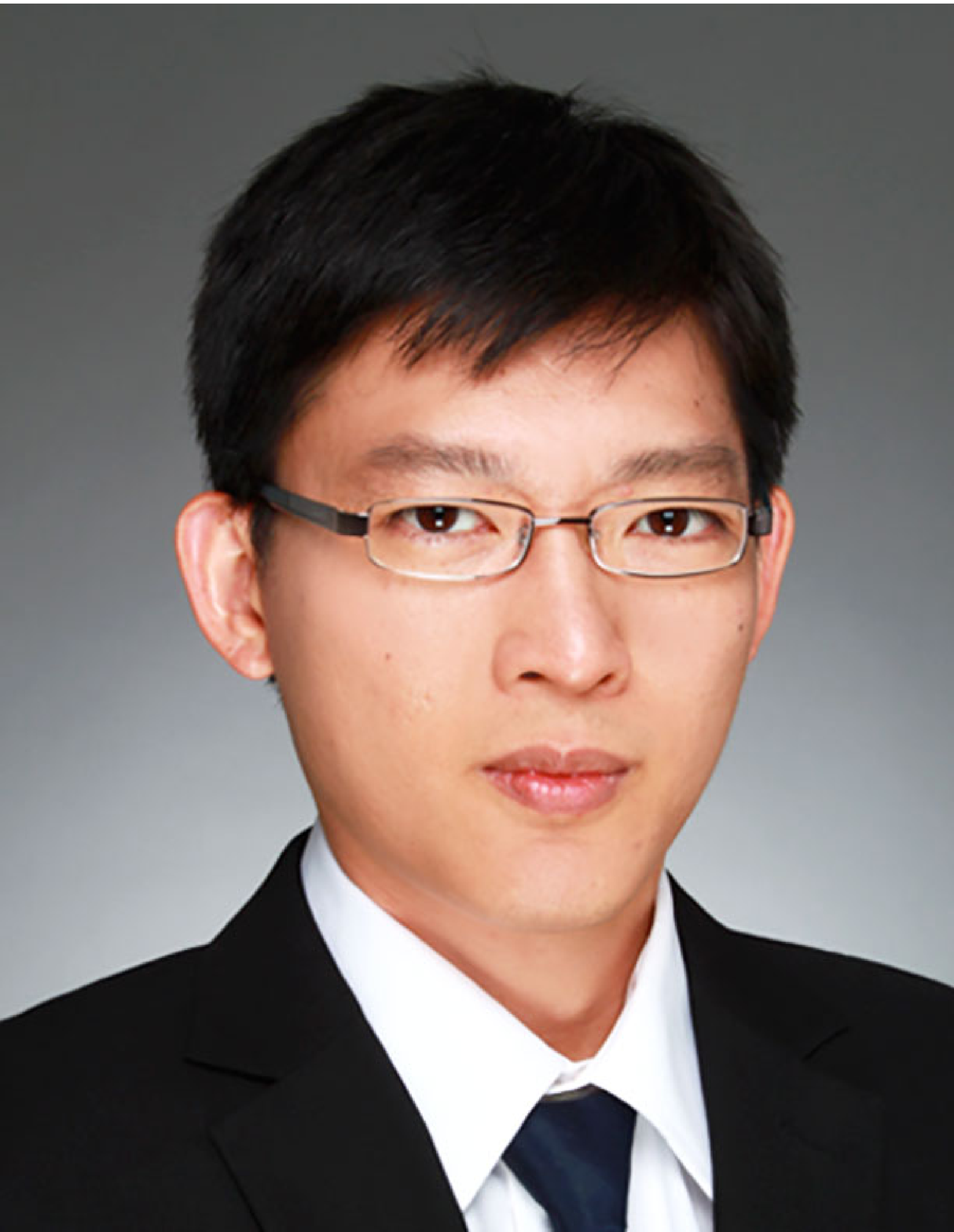}}]{Chin Keong Ho }
	 	(S'05-M'07) received the B.Eng.  (First-Class Hons., minor in business admin.) and
	 	M.Eng. degrees from the Department of Electrical
	 	Engineering, National University of Singapore, in
	 	1999 and 2001, respectively. He received the Ph.D.
	 	degree at the Eindhoven University of Technology,
	 	The Netherlands, where he concurrently conducted
	 	research work in Philips Research. Since August
	 	2000, he has been with Institute for Infocomm Research
	 	 (I$ ^2 $R) , A∗STAR, Singapore. He is Lab Head
	 	of Energy-Aware Communications Lab, Department
	 	of Advanced Communication Technology, in I$ ^2 $R. His research interest includes
	 	green wireless communications with focus on energy-efficient solutions and
	 	with energy harvesting constraints; cooperative and adaptive wireless communications;
	 	and implementation aspects of multi-carrier and multi-antenna
	 	communications. His work in unified study of wireless power and wireless communications received the IEEE Marconi Prize Paper Award in 2015.
	 \end{IEEEbiography}

	 \begin{IEEEbiography}[{\includegraphics[width=1in,height=1.25in,clip,keepaspectratio]{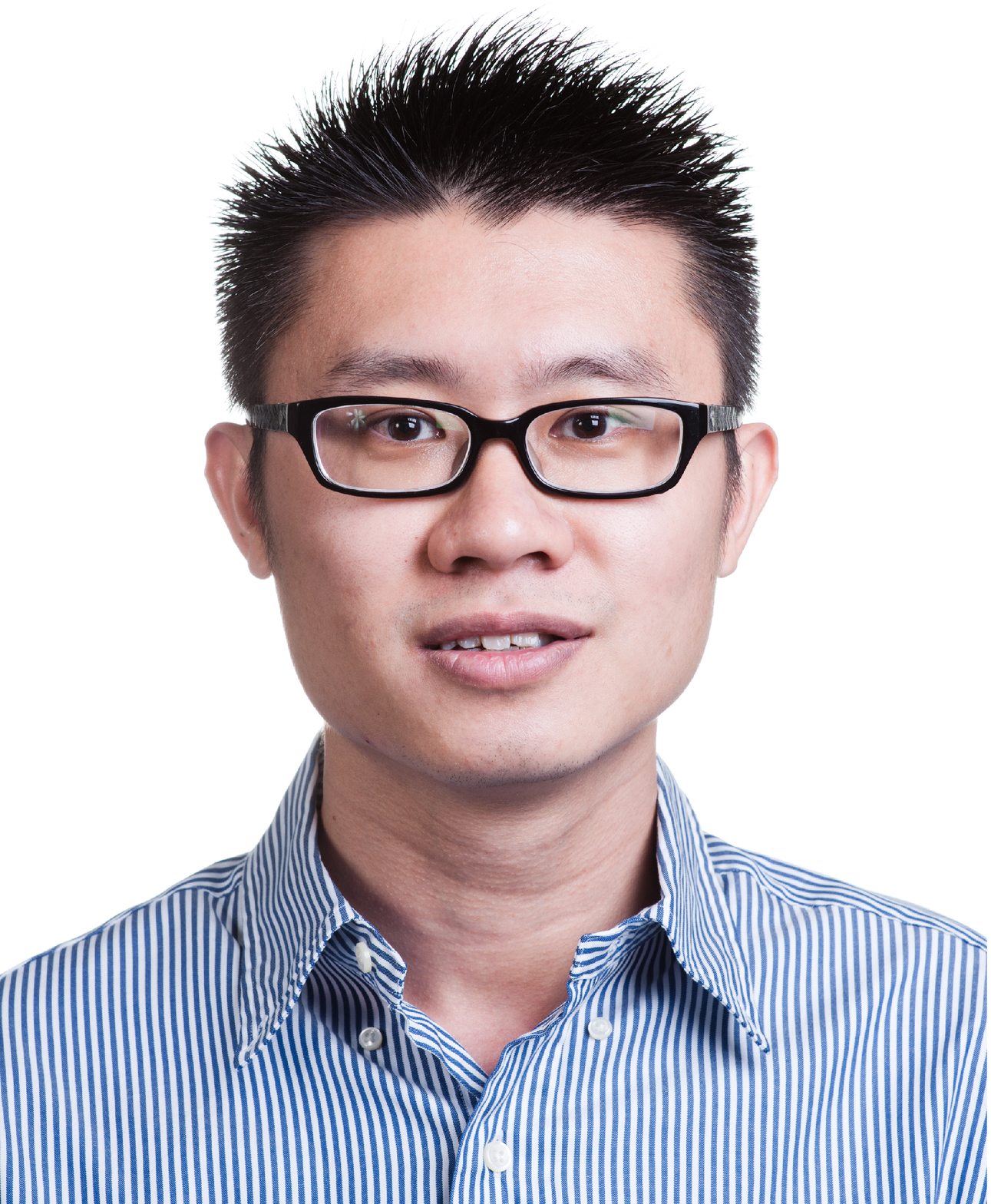}}]{Chau Yuen }
	 	(S'04-M'06-SM'12) received the BEng and PhD degree from Nanyang Technological University (NTU), Singapore, in 2000 and 2004 respectively. Dr Yuen was a Post Doc Fellow in Lucent Technologies Bell Labs, Murray Hill during 2005. During the period of 2006 ‐ 2010, he worked at the Institute for Infocomm Research (I2R, Singapore) as a Senior Research Engineer, where he was involved in an industrial project on developing an 802.11n Wireless LAN system, and participated actively in o Long Term Evolution (LTE) and LTE‐Advanced (LTE‐A) standardization. He joined the Singapore University of Technology and Design from June 2010, and received IEEE Asia-Pacific Outstanding Young Researcher Award on 2012. Dr Yuen serves as an Editor for IEEE Transactions on Communications and IEEE Transactions on Vehicular Technology. He has 2 US patents and published over 300 research papers at international journals or conferences.
	 \end{IEEEbiography}

\end{document}